\documentclass[11pt,a4paper]{article}
\pdfoutput=1
\usepackage{jheppub}

\usepackage{amsmath}
\usepackage{verbatim}
\usepackage{amssymb}
\usepackage{inputenc, array}
\usepackage{textcomp}
\usepackage{appendix}
\usepackage[dvipsnames]{xcolor}
\newcommand{\e}{\epsilon}
\newcommand{\be}[1]{\begin{equation}\label{#1} }
\newcommand{\ee}{\end{equation}}
\newcommand{\bea}[1]{\begin{eqnarray}\label{#1} }
\newcommand{\eea}{\end{eqnarray}}
\newcommand{\p}{\partial}
\newcommand{\refb}[1]{(\ref{#1})}

\renewcommand{\L}{{\mathcal{L}}}

\newcommand{\bL}{\bar{{\mathcal{L}}}}

\renewcommand{\>}{\rangle}

\newcommand{\zc}{|0\rangle_c}

\renewcommand{\th}{\theta}
\newcommand{\g}{\gamma}

\renewcommand{\a}{\alpha}
\newcommand{\ta}{\tilde{\alpha}}

\newcommand{\C}{\tilde{C}}
\renewcommand{\b}{\beta}
\renewcommand{\t}{\tau}
\newcommand{\s}{\sigma}

\newcommand{\js}[1]{\textcolor{red}{\bf [#1 -- js]}}
\newcommand{\mt}[1]{\textrm{\tiny #1}}

\title{A Rindler Road to Carrollian Worldsheets}
\author[a]{Arjun Bagchi,} \author[b]{Aritra Banerjee,}  \author[c]{Shankhadeep Chakrabortty,} \author[a]{Ritankar Chatterjee.}
\author{\\}

\affiliation[a]{Indian Institute of Technology Kanpur, Kanpur 208016, INDIA.\\} 

\affiliation[b]{Okinawa Institute of Science \& Technology, 1919-1 Tancha, Onna-son, Okinawa 904-0495, JAPAN.\\}

\affiliation[c]{Indian Institute of Technology Ropar, Rupnagar, Punjab 140001, INDIA.\\} 

\emailAdd{abagchi@iitk.ac.in, aritra.banerjee@oist.jp, s.chakrabortty@iitrpr.ac.in, ritankar@iitk.ac.in}

\preprint{}

\abstract{The tensionless limit of string theory has recently been formulated in terms of worldsheet Rindler physics. In this paper, by considering closed strings moving in background Rindler spacetimes, we provide a concrete exemplification of this phenomenon. We first show that strings probing the near-horizon region of a generic non-extremal blackhole become tensionless thereby linking a spacetime Carroll limit to a worldsheet Carroll limit. Then, considering strings in $d$-dimensional Rindler spacetime we find a Rindler structure induced on the worldsheet. Novelties, including folds, appear on the closed string worldsheet pertaining to the formation of the worldsheet horizon. The closed string becomes segmented at these folding points and different segments go into the formation of closed strings in the different Rindler wedges. The Bondi-Metzner-Sachs (BMS) or the Conformal Carroll algebra emerges from the closed string Virasoro algebra as the horizon is hit. Quantum states on these accelerated worldsheets are discussed and we show the formation of boundary states from gluing conditions of the different segments of the accelerated closed string.}

\begin{document}
\maketitle

\newpage

\section{Introduction}
Einstein's general theory of relativity is a truly remarkable theory. We had the good fortune of testing the theory to unprecedented accuracies and ``seeing" bizarre and breathtaking objects like black holes \cite{EventHorizonTelescope:2019dse} and ``hearing" the whispers of gravitational waves \cite{LIGOScientific:2016vlm} very recently. The theory also rather spectacularly predicts its own breakdown near spacetime singularities, like the initial Big Bang, or at the centre of black holes. It is expected that a quantum theory of gravity would be able to probe and resolve these singularities of general relativity and this is also why we must construct a theory of quantum gravity. 

\medskip

String theory, over the decades, has emerged as a leading candidate for such a quantum theory of gravity. It is thus very important that one formulates how a string would perceive a singularity in spacetime. In this paper, we build towards this important question. In particular, we delve into details of what happens when a string approaches the horizon of a spacetime singularity. 

\medskip

It is well known that when one goes near the horizon of a spacetime singularity, let's say of a non-extremal black hole to be specific, one encounters a Rindler spacetime. Here we show that a closed bosonic string propagating in Rindler spacetimes has a natural Rindler structure induced on the worldsheet as well. As the string approaches the black hole horizon, it also approaches the horizon of this near-horizon Rindler spacetime. In terms of the worldsheet of the bosonic closed string, in this process of approaching the horizon, the worldsheet acceleration increases and the worldsheet gets more and more deformed from its cylindrical shape. The ultimate process of hitting the horizon means that the string becomes a null string or equivalently a tensionless string. We gave a detailed description of this novel process from the worldsheet in \cite{Bagchi:2020ats}. In this work, we focus on clarifying the role of spacetime and its link to the Rindler physics on the worldsheet described in \cite{Bagchi:2020ats}.   

\begin{figure}[h]\label{fig1}
\centering
  \includegraphics[width=12.5cm]{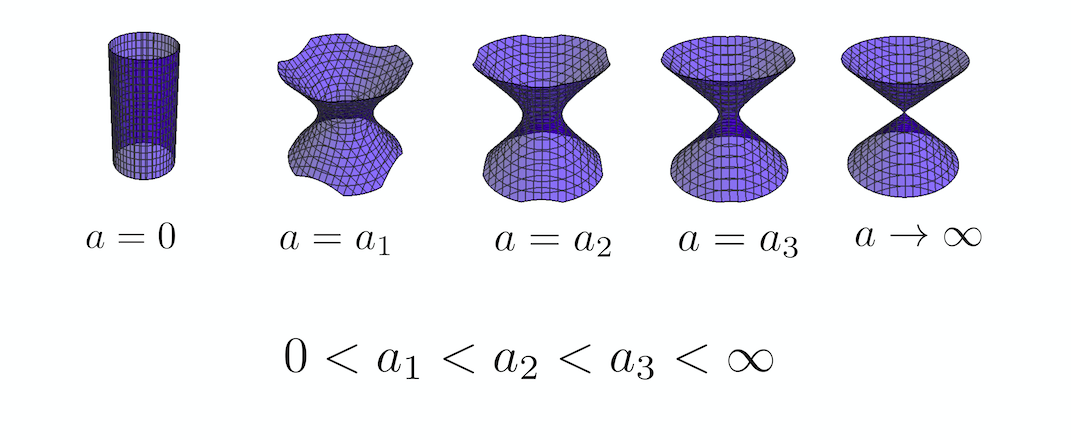}
  \caption{Illustration of how the string worldsheet deforms gradually under increasing acceleration. We represent accelerated worldsheets as hyperboloids, which are a natural generalization of Rindler particle worldlines. }
  \label{}
\end{figure}

As we will go on to describe in the next section, the tensionless limit of string theory gives rise to novel geometric structures on the worldsheet, with Riemannian geometry being replaced by Carrollian geometry. Carrollian geometry is associated with any null manifold and hence has a variety of uses in physics, that is becoming more and more apparent recently. Carrollian physics has emerged in the context of holography in asymptotically flat spacetimes (see e.g. \cite{Bagchi:2010zz, Bagchi:2012cy, Barnich:2012aw, Bagchi:2012yk, Bagchi:2012xr, Barnich:2012xq, Barnich:2012rz, Bagchi:2013lma, Bagchi:2014iea, Hartong:2015xda, Hartong:2015usd, Bagchi:2016bcd} for early work in this direction), where Conformal Carrollian structures have been shown to be isomorphic to the Bondi-Metzner-Sachs (BMS) algebras  \cite{Bagchi:2010zz, Bagchi:2012cy, Duval:2014lpa, Duval:2014uva} which are the asymptotic symmetries of Minkowski spacetime \cite{Bondi:1, PhysRev.128.2851}. Carrollian physics is also finding applications in cosmological scenarios \cite{deBoer:2021jej}, e.g. in the theory of inflation and also in understanding the late time universe. Carrollian structures have been shown to appear in generic black holes \cite{Donnay:2019jiz} and also $2d$ conformal Carrollian symmetry (or equivalently BMS symmetry) has been used for the explanation of black hole entropy in any dimensions \cite{Carlip:2017xne, Carlip:2019dbu} in terms of the BMS-Cardy formula \cite{Bagchi:2012xr}.  

\medskip

In the context of string theory, Carrollian strings have been shown to appear in two different contexts. Firstly, as mentioned above, these degenerate structures appear on the worldsheet when one considers the tension of the string going to zero \cite{Bagchi:2013bga}. Carrollian strings can also appear when the background target spacetime has a natural Carrollian structure \cite{Cardona:2016ytk}. In our paper, we merge these two ideas. We show that when we consider strings nearing a black hole horizon, the string becomes effectively tensionless. The Carrollian structure which arises in the spacetime due to the presence of the black hole horizon thus induces a Carroll structure on the worldsheet which manifests itself in the rescaling of the string tension. 

\medskip

Of course, for a generic non-extremal black hole, there is a $2d$ Rindler metric that appears in the near horizon limit along with a transverse sphere which has a blow-up factor that is associated with it. Thus the near horizon region itself has a degenerate structure, with two Rindler directions scaled down with respect to the spherical directions when one looks at the metric. There is a further Carroll limit when the black hole horizon is hit, which is equivalent to hitting the Rindler horizon in the near-horizon geometry. When considering strings near the horizon of a black hole, the correct thing to do is to quantize string theory in the degenerate near-horizon geometry, which looks very similar to the stringy Newton-Cartan geometry \cite{Andringa_2012} and is possibly a Carrollian cousin of the same. These strings would have a finite tension and the worldsheet would be Riemannian. The metric induced on the worldsheet would possibly be a 2d Rindler metric. The process of hitting the horizon would amount to hitting the worldsheet Rindler horizon and manifest itself by taking the tension to zero. Carrollian structures would appear here and the worldsheet symmetry algebra would change from being two copies of the Virasoro to the BMS algebra. 

\medskip

In this paper, we don't answer this complete question and work with a simpler picture. We consider strings propagating in a Rindler target spacetime and show that a Rindler structure is naturally induced on the worldsheet \footnote{For related literature about strings propagating in Rindler space, one can look at \cite{deVega:1987um, Sanchez:1987yz, Lowe:1994ah} }. In keeping with the expectation above, we show that when the string hits the target space horizon, the worldsheet Rindler horizon is also hit and the string turns null or tensionless. We find many surprising aspects on the worldsheet, including the appearance of foldings, as the closed string accelerates in the background geometry. 
We find intriguing pictures of how the global modes of the accelerated strings are formed by gluing different causally disconnected segments of the strings in different Rindler wedges. We comment on the appearance of open strings from closed strings as the closed strings becomes null, following our earlier observations \cite{Bagchi:2019cay}. We believe that these results would have a direct bearing on the more general question of what happens to strings near black holes and the principal features of our results in this paper would apply to this more general and clearly more involved analysis, which we plan to report on in the near future. 

\medskip

Many of our results seem to have connections with various intriguing observations present in the literature. While admittedly none of these are well examined at this point, we include some comments near the end of the paper with the hope of piquing the interest of the well-versed reader. 

\medskip

The rest of the paper is organised as follows. In Sec \ref{sec2}, we provide a quick recap of the important details of tensionless strings following mainly previous work \cite{Bagchi:2015nca}. In Sec \ref{sec3}, we consider near horizon regions of black holes and show how strings propagating there become effectively tensionless, thus providing a link between Carrollian spacetimes and Carrollian worldsheet theories. From Sec \ref{sec4}, the focus shifts entirely to strings propagating in $d$ dimensional Rindler spacetimes (as opposed to 2$d$ Rindler in the case of black hole near horizons). In Sec \ref{sec4}, we show how the background Rindler induces a Rindler structure on the worldsheet. 
The novelties that this Rindler structure produces on the worldsheet are discussed in Sec \ref{sec5}. This includes the appearance of folds on the worldsheet and a detailed discussion on periodicity. In Sec \ref{sec6}, we discuss how BMS symmetries arise on the worldsheet as the Rindler horizon is approached. Quantum states and Bogoliubov transformations on the worldsheet are addressed in Sec \ref{sec7}. We end with a summary of work in Sec \ref{sec8}, along with a list of comments providing possible links to older literature. 
\medskip

\newpage

\section{Tensionless strings: a quick tour of the basics}\label{sec2}

The tensionless limit of string theory, first studied by Schild \cite{Schild:1976vq}, is the limit where the tension of the fundamental string goes to zero. This is the limit that is opposite to the usual point particle limit, where string theory reduces to Einstein gravity (or its supersymmetric version). This is also the very high energy limit of string theory \cite{Gross:1, Gross:2, Gross:3} and is expected to be of great importance in understanding, amongst other things, the high temperature phase structure and Hagedorn phase transitions \cite{Pisarski:1982cn}.  

\medskip

\noindent We start the discussion of the theory of tensionless strings with the ILST action \cite{Isberg:1993av}, 
\be{LST}
S_{\text{ILST}} = \int d^2 \xi \, V^\a V^\b \p_\a X^\mu \p_\b X^\nu \eta_{\mu \nu}.
\ee
Here $V^\a$ are vector densities, $\eta_{\mu\nu}$ is the background Minkowski metric where the string propagates. To derive this action, one starts with the Polyakov action for the bosonic tensile strings
\be{}
S_p =-\frac{T_0}{2} \int d^2 \xi \, \sqrt{-\g} \g^{\a \b} \p_\a X^\mu \p_\b X^\nu \eta_{\mu \nu},~~~T_0=\frac{1}{2\pi\a'}
\ee
For an explicit tensionless string, the worldsheet metric $\g^{\a \b}$ becomes degenerate and hence has to be replaced by vector densities: $\sqrt{-\g} \g^{\a \b} \to V^\a V^\b$ as $T_0\to 0$. The equations of motion (EOM) of this action $S_{\text{ILST}}$ are given by 
\be{eom}
\p_\a(V^\a V^\b \p_\b X^\mu) = 0, \quad V^\b G_{\a \b} =0, 
\ee
where $G_{\a \b}= \p_\a X^\mu \p_\b X^\nu \eta_{\mu \nu}$  is the induced metric on the worldsheet.

\paragraph{Emergence of BMS:} This worldsheet theory enjoys diffeomorphism invariance similar to its tensile counterpart. We need choose a gauge to fix this redundancy. A particularly convenient choice, analogous to the conformal gauge in the tensile case, is $V^\alpha = (1, 0)$. Again similar to the tensile theory, there is some gauge symmetry left over even after this gauge fixing and the residual symmetry algebra on the  tensionless worldsheet, which replaces the two copies of the Virasoro for the tensile worldsheet, is 
\bea{bms}
&& [L_n, L_m] = (n-m) L_{n+m} + c_L\delta_{n+m,0} (n^3-n), \cr
&& [L_n, M_m] = (n-m)M_{n+m} + c_M\delta_{n+m,0} (n^3-n), \cr 
&& [M_n, M_m]=0.
\eea 
This is the BMS$_3$ algebra \cite{Bondi:1, Sachs:2}, which also arises as the asymptotic symmetries of 3d flat spacetimes at the null boundary.  The natural emergence of BMS$_3$ algebra has been the driving force behind the recent studies of Tensionless strings and its supersymmetric counterparts \cite{Bagchi:2013bga,Bagchi:2015nca,Bagchi:2016yyf, Bagchi:2017cte, Bagchi:2020fpr,Bagchi:2021rfw} \footnote{For other related older literature on null strings, the reader can look at \cite{ Karlhede:1986wb, Lizzi:1986nv, Gamboa:1989px, Gustafsson:1994kr, Lindstrom:2003mg}}. Later in this work we'll show how the classical part of same algebra arises on the worldsheet of an accelerated string in the limit where it hits the Rindler horizon.  

\paragraph{Intrinsic mode expansions:} In the gauge $V^\alpha = (1, 0)$, the EOM for the scalars and the constraints  corresponding to $V^\a$ lead us to:
\be{xeom}
\ddot{X}^\mu=0,\quad \dot{X}\cdot X'=0, \quad \dot{X}^2=0. 
\ee
Now we demand the tensionless worldsheet maintains the closed string boundary conditions of the form $X^\mu(\tau,\sigma)=X^\mu(\tau,\sigma+2\pi)$, like the Tensile one, then the above EOM can solved by the following mode expansion:
\be{mode} 
X^{\mu}(\sigma,\tau)=x^{\mu}+\sqrt{\frac{c'}{2}}B^{\mu}_0\tau+i\sqrt{\frac{c'}{2}}\sum_{n\neq0}\frac{1}{n} \left(A^{\mu}_n-in\tau B^{\mu}_n \right)e^{-in\sigma}. 
\ee
Here the $c'$ is a finite constant included to take care of the dimensions. 
Using this expansion, it is easy to find the form of constraints generating the symmetry algebra
\be{lmab} 
L_n= \frac{1}{2} \sum_{m} A_{- m}\cdot B_{m+n}, \ M_n= \frac{1}{2} \sum_{m} B_{-m}\cdot B_{m+n}. 
\ee
Note here $A,B$ are not the harmonic oscillator modes in the usual sense, and satisfies a different poisson bracket structure,
\bea{AB} 
\{A_m,A_n\}= \{B_m,B_n\}=0; \{A_m,B_n\}= - 2 im\delta_{m+n}.
\eea
 The algebra of the constraints, of course, still leads to the BMS$_3$ algebra as before. To make things more transparent, we need to transform $(A,B)$ into a harmonic oscillator basis: 
\be{CC}
C^{\mu}_n = \frac{1}{2}({A}^{\mu}_n+B^{\mu}_{n}), \quad \C^{\mu}_n =\frac{1}{2}(-{A}^{\mu}_{-n}+B^{\mu}_{-n}).
\ee
It is easy to see that $(C, \C)$ have the usual canonical commutation relations, where $C$ and $\C$ don't talk to each other. One can now also write the mode expansions with respect to these oscillators,
\bea{cexpansion}
X^\mu(\t,\s)&=&x^\mu+2\sqrt{\frac{c'}{2}}C^\mu_0\t +i\sqrt{\frac{c'}{2}}  \sum_{n\neq 0} \frac{1}{n}\left[(C^\mu_n-\C^\mu_{-n})-in\t (C^\mu_n+\C^\mu_{-n})\right]e^{-in\s} \nonumber
\eea
with zero modes given by the momentum,
\be{zero}
C^\mu_0=\C^\mu_0=\sqrt{\frac{c'}{2}}k^\mu.
\ee 
{}The expressions for the classical constraints in this language becomes
\bea{LM} && L_n=\sum_{m} (C_{-m}\cdot C_{m+n}-\C_{-m}\cdot \C_{m-n})  \\ 
&& M_n=\sum_{m} (C_{-m}\cdot C_{m+n}+\C_{-m}\cdot \C_{m-n}+2C_{-m}\cdot \C_{-m-n}). \eea 

As we will go on to see, and as explained in detail in earlier work \cite{Bagchi:2020fpr}, the $C$-oscillators would play an important role in the theory of tensionless strings, effectively defining the quantum structures associated to it. 

\paragraph{Carrollian limit:} The discussion above was aimed at understanding the point in parameter space which corresponds to the tension being exactly equal to zero and hence tensionless strings were treated as fundamental objects. One can recover all of the above results by considering a systematic limiting procedure starting from a theory of tensile closed strings as well. This is a limit where the speed of light on the string worldsheet goes to zero and is called a Carrollian limit. Strings in the tensionless limit can be envisioned become long and ``floppy'' and on top of it all excitations become massless. In terms of coordinates on the string worldsheet, taking this limit entails \cite{Bagchi:2015nca}
\be{URlim}
\s \to \s, \ \t \to \e \t, \ \a'\to c'/\e, \ \e \to 0.
\ee
This contraction pushes the speed of light on this worldsheet to zero and is hence called an Ultra-Relativistic (UR) or a Carrollian limit where the worldsheet becomes a degenerate manifold, with Riemannian structures being replaced by analogues of Newton-Cartan structures, mathematically that of a fibre bundle{\footnote{We wish to clarify that we are still speaking about Carrollian manifolds and not Newton-Cartan manifolds. These are intimately related to each other, in the sense both have fibre bundle structures. Carrollian manifolds have the base and fibre interchanged with respect to the Newton-Cartan manifolds \cite{Duval:2014uoa}. Hence in the rest of the paper, we will sometimes refer to Carrollian structures as ``analogues" of Newton-Cartan structures.}}. If one consistently follows through with the limit at every stage of the computation, it leads to the following contraction of the Virasoro generators
\be{vir2bms}
L_n= \L_n - \bL_{-n}, \ M_n = \e(\L_n + \bL_{-n}),
\ee
where these contracted generators again give rise exactly to the BMS$_3$ algebra. One can now compare the tensile oscillator modes and the tensionless expansions \refb{mode} and derive the simple relation
\be{ABa}
A_n^{\mu} = \frac{1}{\sqrt{\e}} \left( \a_n^\mu - \tilde{\a}_{-n}^\mu \right), \quad B_n^{\mu} = {\sqrt{\e}} \left( \a_n^\mu + \tilde{\a}_{-n}^\mu \right). 
\ee
Here $\a_n$ and $\tilde{\a}_n$ are the tensile oscillators coming from the usual mode expansion,  and these annihilate the tensile vacuum, which we call  $|0\>_{\a}$.
All classical physics of the tensionless string can be reproduced by following the above UR limit. One should bear in mind, the Carrollian limit is usually perceived as a limit on velocities and hence can be thought of exploring the physics of an infinitely boosted worldsheet. 

\medskip

Later in the paper, we will also be interested in the so-called Non-Relativistic (NR) contraction 
\be{NR}
L_n= \L_n + \bL_{n}, \ M_n = \e(\L_n -\bL_{n}),
\ee
In two dimensions this again leads to the classical part of the BMS$_3$ algebra starting from two copies of the Virasoro algebra. Written in terms of co-ordinates on the worldsheet, it is clear this is the opposite limit $\s\to\e\s, \t\to\t$ and is the limit where the speed of light on the string worldsheet goes to infinity.  Hence the name non-relativistic.

\paragraph{Evolving vacua:} Interestingly, switching to the language of  $C$ oscillators, we can define them in terms of the tensile ones using definitions in (\ref{CC}) and (\ref{ABa}). The transformation between $C$ and $\a$ oscillators turn out to be a Bogoliubov transformation, since the canonical structure remains intact:
 \begin{align}
C^{\mu}_{n}&=\frac{1}{2}\Big(\sqrt{\epsilon}+\frac{1}{\sqrt{\epsilon}}\Big)\alpha^{\mu}_{n}+\frac{1}{2}\Big(\sqrt{\epsilon}-\frac{1}{\sqrt{\epsilon}}\Big)\Tilde{\alpha}^{\mu}_{-n} \nonumber \\
\Tilde{C}^{\mu}_{n}&=\frac{1}{2}\Big(\sqrt{\epsilon}-\frac{1}{\sqrt{\epsilon}}\Big)\alpha^{\mu}_{-n}+\frac{1}{2}\Big(\sqrt{\epsilon}+\frac{1}{\sqrt{\epsilon}}\Big)\Tilde{\alpha}^{\mu}_{n},
\label{infvel}
\end{align}
The above relation between oscillators should be valid explicitly near $\e\to0$.\footnote{ For anywhere away from the tensionless regime in the $\e$ parameter space, the vacuum remains the usual tensile vacuum, and only at the extreme limit the change happens.} Surprisingly, as can be seen from the above equations, at $\e=1$ the $C$ oscillators flow to the tensile ones:
\be{ac}
\e=1: \quad  C^{\mu}_{n}=\alpha^{\mu}_{n}, \quad \Tilde{C}^{\mu}_{n}=\Tilde{\alpha}^{\mu}_{n}.
\ee
It is thus natural to envision a flow in the parameter space of $\e$ valid for the whole range $\e \in [0,1]$. The evolving oscillator $C(\e)$ should interpolate between \refb{ac} for $\e=1$ and \refb{infvel} near $\e\to0$. The vacua $|0(\e)\>$ defined for these evolving oscilators $C(\e)$ would also change continuously along the flow and would be given by:
\be{oe}
|0(\e)\>: \, C_n(\e) |0(\e)\> = \C_n(\e) |0(\e)\> = 0, \, \forall n>0.
\ee
At strictly $\e=0$ this will certainly evolve into the purely tensionless string ground state. Note that $|0(\e)\>$ continues to behave like an usual oscillator vacuum throughout the flow. On the other hand, one could look at the evolution of the tensile ground state $|0\>_{\a}$ from the tensionless oscillators' point of view. An inverse transformation of (\ref{infvel}) lead to the tensile vacuum conditions of the form
\bea{bcond}
&& \a^\mu_k|0\rangle_{\a} = (C^\mu_k -\tanh \theta_k\ \tilde{C}^\mu_{-k})|0\rangle_{\a}=0,\   k>0; \nonumber \\ 
&& \tilde{\a}^\mu_k|0\rangle_{\a} = (\tilde{C}^\mu_k-\tanh\theta_k\ C^\mu_{-k})|0\rangle_{\a}=0.
\eea  
Here we have $\tanh\theta=\frac{\e-1}{\e+1}$, which makes sure that (\ref{ac}) is valid at $\e=1$. Note that at the tensionless point $\e=0$, the tensile vacuum turns out to be a special state w.r.t. tensionless oscillators:
\be{za2zc}
|0\rangle_\a = \frac{1}{\mathcal{N'}} \prod_{n=1}^\infty \exp\left[- \frac{1}{n} C_n^\dagger\cdot\tilde{C}_n^\dagger\right]\zc. 
\ee
Here $\mathcal{N'}$ sets the normalization factor. This a Neumann boundary state along all spacetime directions, and in \cite{Bagchi:2019cay} the appearance of this state was interpreted as emergence of open string degrees of freedom from closed string in the tensionless limit.

\medskip

The physics of these tensionless strings as discussed above, is a Carrollian limit on the worldsheet as alluded to earlier, and is essentially an evolution in boosts. Such an evolution does not usually lead to any change in physics. We would only expect boosts to change the physics in the limit where they become infinite. So the Carrollian evolution would only change things like the vacuum structure or lead to changes in the spectrum of the theory when the light-cones close off completely or when we reach $c=0$ on the worldsheet. The above evolution in vacua \refb{oe} can be deemed as a continuous evolution and hence clearly not an evolution in boosts. As it turns out, we can explain this particular evolution in parameter space very naturally by accelerating string worldsheets. The reason why the Carrollian limit works is that the limit of infinite boosts and the limit of infinite accelerations would land one up on the same identification \refb{infvel}. It turns out that both scenarios take us extremely close to the horizon of the underlying Rindler spacetime, which we will elaborate more in this work.

\newpage

\section{Strings near Black Holes} \label{sec3}
Tensionless strings can arise in several cases, e.g. in the very high energy sector of string theory where string scattering amplitudes simplify \cite{Gross:1}, near the Hagedorn phase transition of string theory \cite{Pisarski:1982cn, Olesen:1985ej,Atick:1988si}, in the worldsheet formulation of ambitwistor strings \cite{Casali:2016atr}. In this paper, we would be interested in understanding how strings behave near spacetime singularities. 
This has been an oft-asked question in the literature since the early days of string theory, and several schools of thought exist. In this paper, we shall show explicitly how this is related to the tensionless limit of string theory. 
In particular, and in this section, we will show that that for strings propagating in black hole spacetimes, as the string nears the black hole event horizon, the tension of the string decreases. The near horizon limit of a  black hole geometry thus can be understood as a tensionless limit on a worldsheet.{\footnote{The calculations in this section were carried out in collaboration with Joan Simon.}} 

\subsection{Rindler backgrounds near black hole horizons}
Before we move into the details of our particular computations, we remind the reader of some rudimentary basics and show how Rindler spacetimes arise in the near-horizon limit of black hole geometries. We start with the trusted example of a non-extremal Schwarzschild black hole in four dimensions, whose metric reads 
\be{schw}
ds^2 = -\left( 1 - \frac{2M}{r}  \right)dt^2+\left( 1 -  \frac{2M}{r} \right)^{-1}dr^2+r^2 d\Omega_{2}^2,
\ee
where $2M$ is again the radius of the horizon ($M$ is the mass of the black hole, and we use units where Newton's constant $G=1$) and we use the two dimensional sphere as transverse coordinates to simplify notation. To take the near horizon limit on this metric, one makes the substitution $r\to 2M+\frac{u^2}{2M}$ and approximates $\frac{u}{2M}\ll 1$. Zooming in near the horizon, the above metric becomes
\be{schwrind}
ds^2 = -\left(\frac{u}{2M} \right)^2 dt^2+4 du^2+(2M)^2 d\Omega_{2}^2.
\ee
Note that all the corrections to this metric are of $\mathcal{O}\left(\frac{1}{2M}\right)$. Now this metric has clear Rindler structure, and since the geometry factorises into $R^{1,1}\times S^2$, we can take the $(u,t)$ space and perform a simple coordinate transformation,
\be{rindcoord}
e^{a X} = 2u,~~~\mathcal{T} = \frac{t}{4a M},
\ee
where the $a$ is the acceleration parameter by definition. Now we could see the $(1+1)$ dimensional near horizon spacetime becomes a Rindler spacetime in familiar coordinates,
\be{}
ds_2^2 = a^2 e^{2a X}\left(-d\mathcal{T}^2+dX^2 \right).
\ee

\subsection{Background near horizon limit as a Carrollian limit }
We will now briefly recall the identification of the near horizon limit as a Carrollian (or Ultra-Relativistic, $c\to 0$) limit following \cite{Donnay:2019jiz}, an idea central to studying the symmetries at the horizon. Again, for simplicity, we stick to the four-dimensional Schwarzschild black hole. The metric, now in Eddington-Finkelstein (EF) coordinates $(v,r,\theta^{i})$ is given by
\begin{equation}\label{schwa}
    ds^2=-\Big(1-\frac{2M}{r}\Big)dv^2+2dvdr+r^2h_{ij}d\theta^{i}d\theta^{j}
\end{equation}
where $v=t+r^*$. Near a smooth null Gaussian hypersurface, Gaussian null coordinates ($v, \rho, \theta^i$) can be introduced where a generic near horizon metric takes the following form
\begin{equation}
    ds^2=-\phi(v,\theta^k) \rho dv^2+2dvd\rho+2 h_i(v,\theta^k) \rho d\theta^idv+h_{ij}(v,\theta^k)d\theta^id\theta^j.
\end{equation}
In EF coordinates, near the horizon ($r\to 2M$), the metric (\ref{schwa}) can be shown to become 
\begin{equation}
    ds^2=-\frac{\Theta}{2M}dv^2+2dvd\Theta+(2M)^2h_{ij}d\theta^{i}d\theta^{j}
\end{equation}
where $\Theta=r-2M$. 
 This clearly is in null Gaussian coordinate form with $h_a=0$, $h_{ab}=(2M)^2h_{ij}$, $\phi=\frac{1}{2M}$. In this setup constant $\Theta$ hypersurfaces ($\Theta=k$) have the following induced metric
\begin{equation}\label{nhsc}
    g_{\alpha\beta}=\begin{bmatrix}
    -\frac{k}{2M} & 0\\0 & (2M)^2h_{ij}
    \end{bmatrix}
\end{equation}
Now one recalls that any spacetime can be written in the so-called Randers-Papapetrou parametrization, where the general form of a metric is given by 
\begin{equation}
ds^2= -\Omega c^2dv^2+2c^2\Omega b_id\theta^idv+(\Omega_{ij}-c^2b_ib_j)d\theta^id\theta^j, 
\end{equation}
where $\Omega$, $\Omega_{ij}$, and $b_{i}$ are functions of coordinates ($v,\theta^i$).
From this, it is clear that \refb{nhsc} is in Randers-Papapetrou form with the following identifications
\be{}
\Theta=k=c^2, \quad b_{i}=0, \quad \Omega=\frac{1}{2M}, \quad \Omega_{ij}=(2M)^2h_{ij}.
\ee 
It is thus evident that the near horizon limit $\Theta\to0$ is a Carrollian limit $c\to0$. In this limit the induced metric $g_{\a\b}$ clearly becomes degenerate. 

\medskip

\subsection{Near Horizon limit as Tensionless limit on worldsheet}

Now that we know that spacetime near horizon limits are Carrollian in nature, we ask our main question, what happens to a string worldsheet as the string probes a black hole and moves into its near horizon geometry?
We begin with strings propagating on a black hole background and take the near-horizon limit directly to show that tensionless strings emerge as we go near the horizon. Again for simplicity, we will begin with a Schwarzschild metric in four dimensions with the sphere written out in terms of coordinates $\theta^k$ (generalizations to higher dimensions and/or other geometries  is immediate): 
\be{}
ds^2= - \left(1-\frac{2M}{r}\right)dt^2 +  \left(1-\frac{2M}{r}\right)^{-1}dr^2 + r^2 h_{ij} (\th) d\th^i d\th^j.
\ee
For the present purpose it is well suited to introduce tortoise coordinates $r_*$: 
\be{}
r_*= r + 2M \log{\left(r-2M\right)}
\ee
to rewrite the metric as
\be{sch-tor}
ds^2= \left(1-\frac{2M}{r}\right) (-dt^2+dr_*^2) + r^2(r_*) h_{ij} d\th^i d\th^j.
\ee
We now define the near horizon region of the spacetime as
\be{}
r= 2M(1+\rho), \quad 0<\rho\ll 1.
\ee
In terms of the tortoise coordinates, we could have the identification:
\be{}
\rho \sim \frac{\mu}{2M} e^{r_*/2M}, \quad r_*\to -\infty
\ee
where the dimensionless parameter $\mu$ has been introduced to lend the limit a further handle. One can also introduce  a critical radius $r_c$, where, 
\be{}
r_*=2M(r_c+\delta r)
\ee
so that the near horizon limit effectively becomes $r_c\to -\infty$. We now look at strings propagating in this geometry. The Polyakov action is 
\be{}
S =-\frac{T_0}{2} \int d^2 \xi \, \sqrt{-\g} \g^{\a \b} \p_\a X^\mu \p_\b X^\nu g_{\mu \nu}(X)
\ee
where $g_{\mu \nu}$ is the background Schwarzschild metric \refb{sch-tor}. The induced metric on the worldsheet is thus given by
\bea{indmet1}
G_{\a\b} &=& \p_\a X^\mu \p_\b X^\nu g_{\mu \nu}(X) \nonumber \\ &=& \left(1-\frac{2M}{r}\right)\left(- \p_\a t \p_\b t + \p_\a r_* \p_\b r_* \right) + r^2(r_*) h_{ij} \p_\a \th^i \p_\b \th^j. 
\eea
Now we examine the fate of the above metric in near horizon limit. In this limit, in our  parameterisation, for the induced metric we get 
\bea{}
G_{\a\b} &\approx& \frac{\mu}{2M} e^{r_c} e^{\delta r} \left\{- \p_\a t \p_\b t + (2M)^2 \p_\a (\delta r)  \p_\b (\delta r) \right\} + r^2(r_*) h_{ij} \p_\a \th^i \p_\b \th^j. \\
&=& \frac{\mu e^{r_c}}{2M} \big[ e^{\delta r} \left\{- \p_\a t \p_\b t + (2M)^2 \p_\a (\delta r)  \p_\b (\delta r) \right\} + (2M)^2 \left\{ \frac{2M}{\mu} + 2 e^{r_c} e^{\delta r}\right\} e^{- r_c} h_{ij} \p_\a \th^i \p_\b \th^j \big]. \nonumber
\eea
Notice here that we need to take a decompactifying limit on the transverse sphere in this near horizon limit, replacing the 2-sphere by a 2d plane{\footnote{This is an approximation. If we wanted to keep the sphere, we would need to work in the analogue of a Newton-Cartan geometry and consider strings propagating there, as we have mentioned in the introduction. This more involved calculation will be reported in upcoming work.}}. Putting this back into the Polyakov action and keeping the worldsheet metric unchanged, we see that in this parametrization, we are left with a rescaled string tension $T$: 
\be{}
T = \frac{\mu}{2M} e^{r_c}  \, T_0.
\ee
The near horizon limit $r_c\to -\infty$ is thus a tensionless limit on the string worldsheet. In terms of our analysis of tensionless strings in the previous section, this means the identification
\be{}
\e = \frac{\mu}{2M} e^{r_c},
\ee
 where $\e$ is the contraction parameter defined in (\ref{URlim}).

\medskip

So the takeaway from the above is that the near horizon limit induces a limit on the strings propagating in a black hole background where the tension of the string goes to zero. In the previous section, we have also argued how the direct tensionless limit (\ref{URlim}) can be understood as an ultrarelativistic or Carrollian limit on the string worldsheet itself. Given that we also learned the near horizon limit is indeed a Carrollian limit on the background spacetime, the synthesis of this section is now clear: \textit{Carrollian limit in the background spacetime induces a Carrollian limit on the string worldsheet}.

\newpage

\section{From Rindler spacetimes to Rindler worldsheets}\label{sec4}

In previous work \cite{Bagchi:2020ats}, we had formulated the tensionless limit as a limit of increasing worldsheet accelerations. This gave rise to Rindler physics on the worldsheet, with the null string emerging when the Rindler horizon was hit. The crucial observation was that the inertial $\a$ oscillators were linked to the intrinsic $C$ oscillators by a worldsheet Bogolioubov transformation \refb{infvel} and hence we attempted to build a family of Bogoliubov transformations that would continuously link the tensile worldsheet with the tensionless one, as acceleration was dialled from zero to infinity. Near the horizon, or at very very high acceleration, the form of these transformations would mimic \refb{infvel}.  

\medskip

As we have stated previously, strings become tensionless in several cases, and in the last section we learnt that they appear when a worldsheet moves into the near horizon region of a black hole. We also learnt that the near horizon region of a black hole spacetime looks locally like a Rindler one.
\medskip

Motivated by this, in this paper, we are interested in a particular application of this general procedure to study accelerated worldsheets. As mentioned in the introduction, we would be interested in strings traveling in a {\em background Rindler spacetime}. We will argue how this induces a Rindler structure on the worldsheet, thereby providing a concrete realisation of our worldsheet methods in \cite{Bagchi:2020ats}. 

\medskip

We will then show various aspects of worldsheet physics emerging because of the embedding into the background . These would match with previous considerations made purely from the point of view of the worldsheet and hence provide a different yet complimentary perspective of novel physics pointed out before. In particular, building on material we present here, we will show in the next section the emergence of Bogolioubov transformations on the worldsheet, now motivated from the spacetime. 

\medskip
We add a disclaimer here. The analysis of the paper from this section onwards is going to be in a $d$-dimensional background Rindler spacetime and {\em{not}} the near-horizon of a generic black hole, which contains a $2d$ Rindler times transverse directions. We will, for the moment, circumvent any potential issues with the transverse directions and return to this in later work. However, we expect the main features of our analysis in this paper to be valid for the analysis of strings in the near horizon of black holes.

\subsection*{Background Rindler and worldsheet Rindler}
We now move on to our study of strings moving in Rindler spacetimes or their analogues.
Let us start by reminding ourselves some basics. The Polyakov action for tensile bosonic strings moving in a $D$-dimensional spacetime is given by
\begin{equation}
    S=\frac{1}{4\pi\alpha'}\int d\sigma d\tau \sqrt{\g}\g^{\alpha\beta}g_{AB}(X)\partial_{\alpha}X^{A}\partial_{\beta}X^{B}.
\end{equation}
Here $g_{AB}$ and $\g_{\alpha\beta}$ are spacetime and worldsheet metrics respectively, $X$ is the worldsheet scalar and $T=\frac{1}{2\pi\alpha'}$ is the string tension. The equation of motion and the Virasoro constraints associated with the above action are
\begin{align}
    &\Box X^{A}+\Gamma^A_{BC}(X)\partial_{\alpha}X^{A}\partial_{\beta}X^{B}=0\\
    &T_{\alpha\beta}=g_{AB}(X)\partial_{\alpha}X^{A}\partial_{\beta}X^{B}-\frac{1}{2}\g_{\alpha\beta}g_{AB}(X)\partial_{\sigma}X^{A}\partial^{\sigma}X^{B}=0.
\end{align}
Here $\Box$ is the D'Alembertian of the worldsheet.  The action is invariant under reparametrizations of the worldsheet coordinates given by
\be{}
    \delta X^{\beta}=\epsilon^{\alpha}\partial_{\alpha}X^{\beta}, \quad
    \delta \g_{\alpha\beta}=\epsilon^{\rho}\partial_{\rho}\g_{\alpha\beta}+(\partial_{\alpha}\epsilon^{\rho})\g_{\rho\beta}+(\partial_{\beta}\epsilon^{\rho})\g_{\alpha\rho}
\ee
allowing us to choose $\gamma_{\alpha\beta}$ to be in conformally flat i.e.
\begin{equation}
    \g_{\alpha\beta}=\Lambda(\sigma, \tau)\eta_{\alpha\beta}.
\end{equation}

\begin{figure}[h]
\centering
  \includegraphics[width=7.5cm]{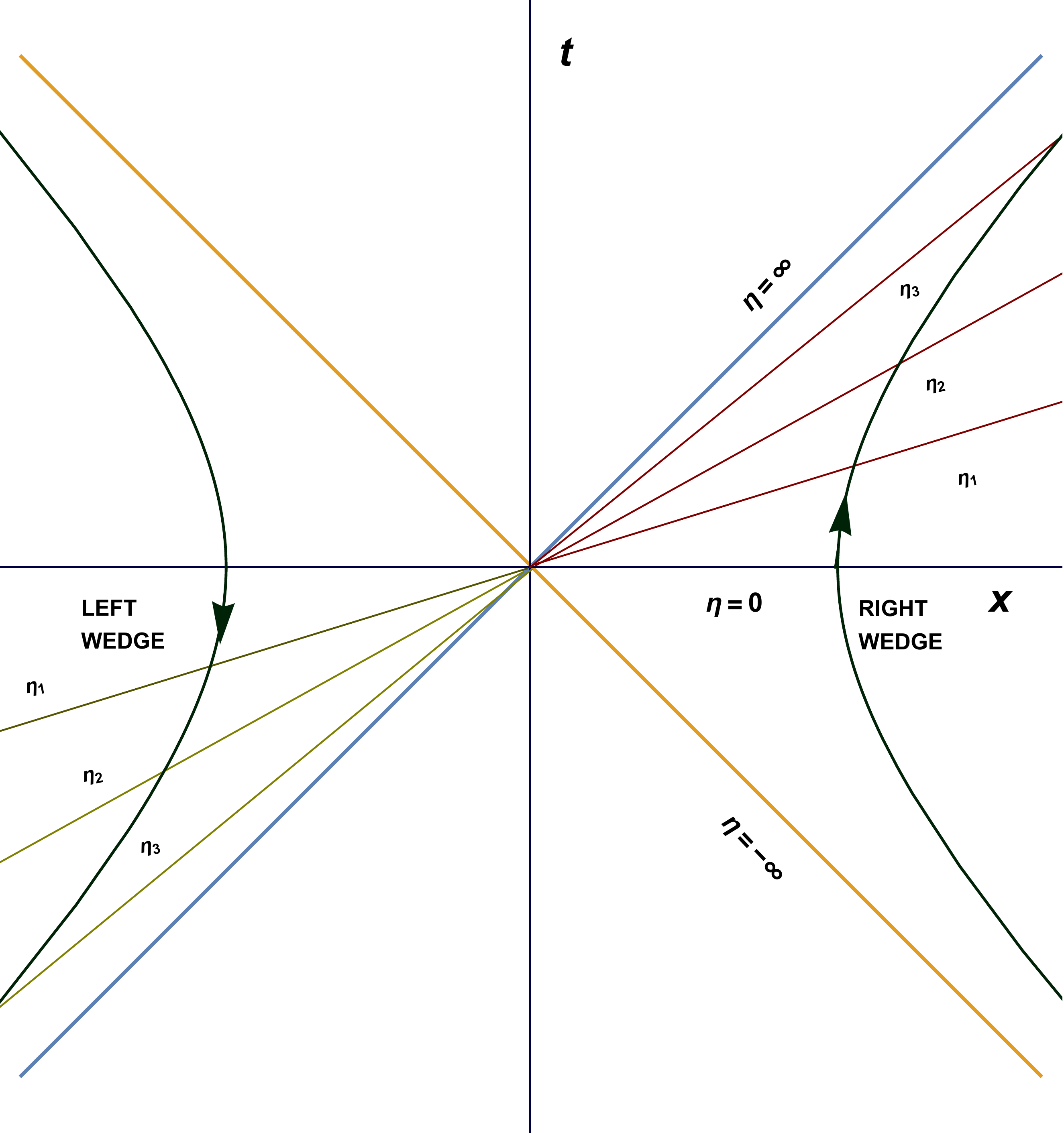}
  \caption{Two dimensional Rindler space spanned by $(\eta,\xi)$ coordinates, with the inertial set of coordinates being $(t,x)$. The timelike vector flows in opposite directions for two causally disconnected wedges (Right/Left), and the constant $\eta$ timeslices $\eta_1<\eta_2<\eta_3$ are shown spanning both wedges. Constant $\xi$ surfaces are families of hyperbolae corresponding to progressively accelerated observers. Light cones in Minkowski $(t,x)$ coordinates become Rindler horizons for accelerated observers. }
  \label{rsp}
\end{figure}
\medskip

\medskip

Now we consider a string worldsheet moving in a Rindler spacetime, which in turn is given by a transformation on the inertial $d$ dimensional Minkowski background {\footnote{Note this is the map for the right Rindler wedge in spacetime. For the sake of completeness, we briefly present the structure of two-dimensional transformed part of the Rindler space in Figure (\ref{rsp}).}}:
\be{}
    X^{0}=e^{a X'^{1}}\cosh{(a X'^{0})}, \quad X^{1} =e^{a X'^{1}}\sinh{(a X'^{0})}, ~X^{i} = X'^{i}.
\ee
Here the primed coordinates are that of the accelerated frame with an acceleration $a$ (along $X^1$), and the unprimed ones are that of the inertial frame. The metric of this new target spacetime is
\begin{equation}
    ds^2=a^{2}e^{2a X'^{1}}[-(dX'^{0})^{2}+(dX'^{1})^{2}]+(dX^{i})^2.
\end{equation}
Following our discussion earlier, one could consider our probe string having a worldsheet moving into a near horizon black hole background as opposed to a flat one. Here the $X^0$ and $X^1$ target space directions in the inertial spacetime  are transformed to the Rindler coordinates and other directions(denoted by $i$) stay the same.  Rindler horizons here correspond to the regimes where $X'^{1}\to -\infty$ and $X'^{0}\to \infty$.

\medskip

So now we have a notion of two distinct observers associated to two worldsheets, one inertial observer, who moves in a flat space, and another accelerated observer, who moves in a Rindler spacetime. 
We will consider a closed string propagating in this Rindler background. The induced metric on the worldsheet has a form
\be{accmet}
G_{ab} \propto a^2 e^{2a X'^{1}}\p_a X^\mu \p_b X^\nu \eta_{\mu\nu}.
\ee
For the moment we bypass the issue of the transverse space and consider further that the $i$ directions are not functions of worldsheet coordinates, then the string effectively moves in two dimensions. This induced metric has a curious structure in the sense that this vanishes in the limit $X'^{1}\to -\infty$, i.e. when the target space horizon is hit. This degeneration of the induced metric can be seen as the worldsheet becoming null. 

\medskip

Away from the point where the string hits $X'^{1}\to -\infty$, everything is regular on the worldsheet and in general one can use Weyl symmetries in two dimensions to do away with the conformal factor in (\ref{accmet}) so that worldsheet conformal gauge remains intact. Only near $X'^{1}\to -\infty$, the Weyl factor becomes non-invertible and one can't perceive a worldsheet flat metric anymore as it degenerates. Remember this limit also means taking the target space acceleration to infinity in Rindler coordinates and hence we can say that \textit{null strings emerge in the limit of infinite Rindler acceleration}, a proposal that we'll be elaborating in the later sections. This goes well with our discussion in the previous section where the null string would emerge as a worldsheet nears a black hole horizon.

\medskip

Now on the worldsheet in question, the target space coordinates are embedded through pullback functions of worldsheet coordinates. If we call the new non-inertial worldsheet coordinates $(\theta, \psi)$, the values of the embedding functions touch $X'^{1}\to -\infty$ and $X'^{0}\to \infty$ at certain sets of values for these worldsheet coordinates, regardless of happenings in the target space. For example if we take 
\be{} X'^{0} = \theta \quad X'^{1} =f(\theta, \psi),
\ee 
then by pullback  $X'^{1}\to -\infty$ would mean the embedding function encodes a divergence on the worldsheet for certain values of the coordinates i.e. 
\be{}
f|_{\psi = \psi_*}\to -\infty.
\ee 
Looking back at the form of (\ref{accmet}), these special points will be always present on the non-inertial worldsheet \footnote{This is especially clear if there is still some periodicity in $\psi$, the spatial coordinate. In that case there will always be points in a full period of $\psi$ where the pullback map will diverge, making sure those points are present even when other parts of the worldsheet are at a finite acceleration.} and the metric will degenerate at these points. As we mentioned earlier, these are the worldsheet points where the Weyl undoing of the conformal factor fails, and we designate these as the points where a worldsheet horizon has been induced from the target space. In the next section, we will interpret these special points as `folds' on the worldsheet. We stress that when a worldsheet completely `falls' into the horizon, all the points on the string correspond to these diverging embeddings, and the whole string becomes null.

\medskip


One should note here that although we have arrived at our logic using a very simple argument, it can be generically understood that the induced metric on a worldsheet near a spacetime horizon will always take a form where an analog of the Rindler horizon will be clearly present. For example in the case of (\ref{indmet1}), if we disregard the transverse coordinates, we can see it clearly has a two dimensional Rindler geometry on the worldsheet with $e^{2af}\sim 1-\frac{2m}{r}$, with touching the Schwarzschild horizon ($r=2m$) signifying all points on the worldsheet becoming degenerate.
\medskip

So in this simplistic set-up, we have been able to show that when a string moves in a Rindler spacetime, the background will induce a Rindler structure on the worldsheet of the string as well. In particular, \textit{the presence of background Rindler horizon induces a worldsheet Rindler horizon}.  Let us emphasize that the above is a universal feature that we expect to hold in any worldsheet embedding as long as the embedding functions have the right analytic structures. The particular embedding chosen earlier is for demonstrating these features in a simplified set up. 

\medskip

%

We can envision an extreme scenario as well, where as the string almost strikes (or intersects) the target space horizon and the worldsheet is very constricted in the sense the spacetime Rindler horizon and the worldsheet horizon has to coincide. In this situation we can again easily see there will be particular points on the worldsheet where the embedding diverges, and this divides the string into causally disconnected segments. In all settings, the emergence of the full null string will be visualized by something like the last picture of Figure (\ref{fig1}) where the whole string lies along the Rindler horizon and all points on the worldsheet are degenerate. 

\medskip
This coincident position is a precarious situation, and one might think this as the spatial embedding function $X'^{1}$ becoming linear in worldsheet spatial coordinates as well, e.g. 
\be{} X'^{0 }= \theta, \ X'^{1} =\psi_*+c
\ee 
where $c$ is a constant and $(\theta,\psi)$ are the coordinates on our new ``accelerated'' worldsheet. The induced metric in this case becomes,
\be{indmet}
ds^2 =a^2 e^{2a (\psi_*+c)}(- d\theta^2+ d\psi_*^2).
\ee
Although this example again looks like a worldsheet Rindler metric \footnote{However $c$ could be a very large positive constant which effectively regulates the effect of $\psi_* \to -\infty$ and stops the metric from becoming null.}, as readers may speculate, in this extereme limit it is hard to imagine the string as a proper closed string. This expectation would be borne out by our analysis in later sections when we find an open string emerging from closed strings in the extreme limit. This is what was also argued in \cite{Bagchi:2019cay}.

\medskip

Independently, we can discuss similar change of physics just from the worldsheet point of view, and that should coincide with the discourse of induced Rindler structures above. 
It is very clear that we are still allowed to do the following reperametrization on inertial worldsheet coordinates so that metric corresponding to the one on accelerated worldsheet takes the form of a two dimensional Rindler space,
\be{st}
    \sigma =e^{a\xi}\cosh{a\eta}, \quad \tau =e^{a\xi}\sinh{a\eta},
\ee
 with which the $2d$ metric takes the textbook form
\be{rindm}
ds^2 =a^2 e^{2a \xi}(- d\eta^2+ d\xi^2).
\ee  
We will take this as an intrinsic mapping between our inertial and accelerated worldsheets, and it can be considered exactly equivalent to the structure obtained by our previous considerations albeit agnostic of the happenings in the target space. The inverse transformation is given by 
\be{inverse}
    \xi=\frac{1}{2a}\log{[\sigma^2-\tau^2]}, \quad \eta=\frac{1}{2a}\log{\Big[\frac{\sigma+\tau}{\sigma-\tau}\Big]}.
\ee
This map is clearly divergent at $\s = \pm\t$, and corresponds to the points where the conformal factor in (\ref{rindm}) vanishes . So there are at least two singular points present on each constant timeslice of the non-inertial worldsheet which correspond to perfectly regular points on the inertial one. Our argument in this section was these points signify appearance of  worldsheet event horizons and in the upcoming section, we will argue that the string folds along these points. Nonetheless, in what follows we will call the string with $(\eta,\xi)$ worldsheet coordinates as the Rindler/ Accelerated string, which is different from the inertial string with $(\t,\s)$ worldsheet parameterization.

\newpage

\section{Novelties of the Rindler worldsheet}\label{sec5}
Keeping in mind the particular connection to the embedding in a Rindler spacetime we described in the previous section, we will now see various curiosities of the Rindler worldsheet. It is important to remember here that this is not a pure worldsheet perspective, but features which appear when the closed string moves on a Rindler background and the embedding, or the knowledge of the background, plays the central role in our understanding here. 

\subsection{Appearance of folds: from spin to acceleration}

We first delve into the appearance of folds on the worldsheet. We begin with some background about folded strings before we discuss a novel folding mechanism for our specific case. It is well know that flat space rotating strings have asymptotically linear Regge trajectories \footnote{String regge trajectories are plotted as $J$ as functions of $E^2$.}, i.e. the relation between energy ($E$) and angular momentum ($J$) is given by \cite{Green:1987sp}:
\be{}
J= \alpha' E^2 +\mathcal{O}(E^{3/2}).
\ee
The $\mathcal{O}(E^{3/2})$ correction is interesting in the sense that one can get this in a theory of open strings with massive endpoints. In a theory of pure closed strings this correction will come from `folding' points on the string. These points appear when the string is divided into two segments with gluing boundary conditions at two ends. This particular scaling of conserved quantities happens only in flat space. In contrast in AdS spacetimes, one would have a logarithmic scaling with spin for folded strings in the high-spin limit \cite{Gubser_1998}. 

\smallskip

A fold is a point where the map from target space to worldsheet coordinates diverges.
The determinant of the induced two dimensional metric vanishes at the folding points. Hence the  worldsheet curvature diverges and this is not just a coordinate singularity. Let us consider the case of static gauge strings, where the embedding is $X^0 = \t$ and $X^1 = X(\t,\s)$, and we use the remaining reparameterisation in $\s$ to put cross terms on induced metric to zero, with other target space directions independent of worldsheet coordinates. The induced metric in this case is written as 
\be{}
\begin{pmatrix}
-1+ (\p_\t X)^2 & 0 \\
0 & (\p_\s X)^2
\end{pmatrix}
\ee
Considering $\p_\tau X = \beta$ is the proper velocity on any point on the worldsheet, we find the determinant of the induced metric
\be{}
\text{det}(h_{\a\b}) = (1-\beta^2)(\p_\s X)^2,
\ee
so the folds can appear on points moving at the velocity of light i.e. where $\beta \to 1$, or in other words, the points where the worldsheet metric degenerates. As mentioned earlier, these endpoints will have pure open string boundary conditions and there will be no reflection at these boundaries. The total string however remains closed via gluing conditions imposed on these particular points, which act as boundary conditions in defining the string profile. The idea of folded `closed' strings being made up of `open' segments glued together will be a crucial idea in this work.

\begin{figure}[t]
\centering
  \includegraphics[width=12.5cm]{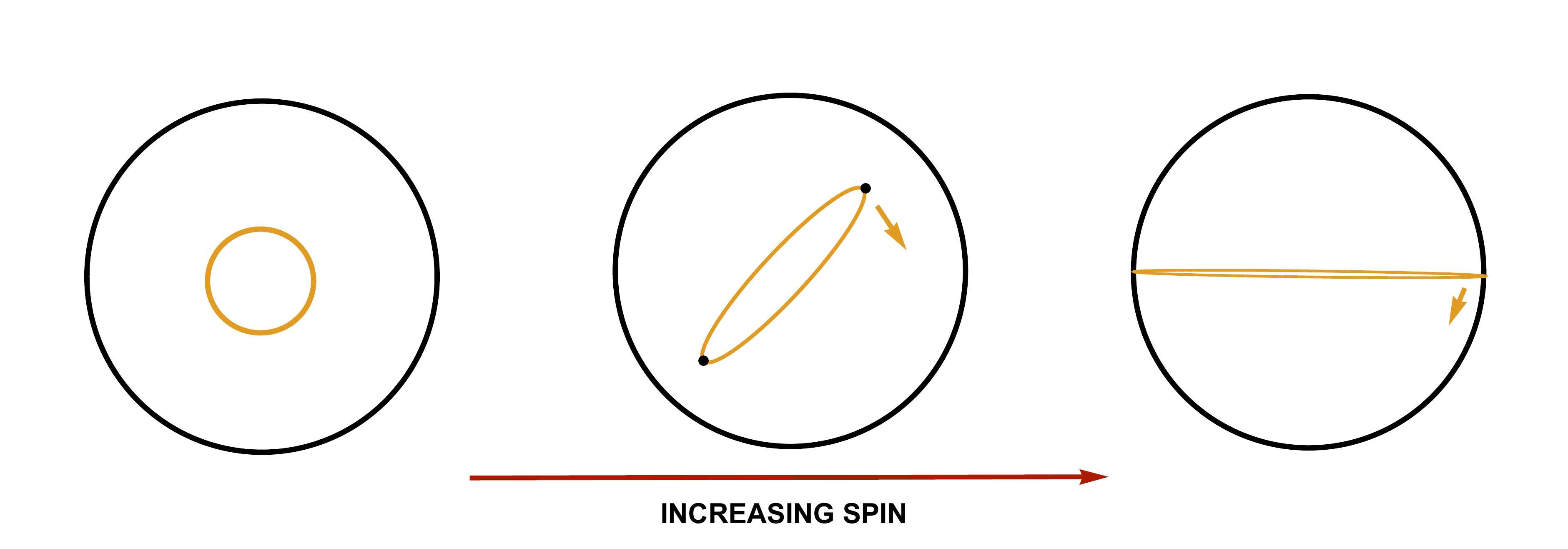}
  \caption{ Illustration of a spinning folded string in the AdS cylinder. At the small spin limit the string is circular and stays near the center, as spin is increased folds start to appear at marked points on the worldsheet, and finally the folding points touch the boundary. }
  \label{spinning}
\end{figure}

\medskip

A similar phenomenon occurs in spinning strings in AdS \cite{Gubser_1998}, which have been studied extensively over the years due to implications for $AdS/CFT$ correspondence. The spinning string in AdS exists in two distinct phases, the ``short string'' limit and the ``long string'' limit, corresponding to small and large spin values $j$, see figure (\ref{spinning}) for a description. The energy-spin dispersion relation varies widely in these two limits. While in the small spin limit the string behaves more like a circular string staying near the AdS centre, as one increases spin, folds (or ``cusps'') start to appear on the string as it increases very fast in length towards the boundary.  The scale of the spin value is set by the string tension, since the contracting effect due to the tension is pitched against the spin for the string. To be more precise strings with $ j \ll T$ behave like circular strings with flat space regge behaviour while those with $j\gg T$ are very long folded strings with logarithmic energy-spin relation.  \footnote{In the $AdS/CFT$ context one uses the t' Hooft coupling $\lambda$ to define $\mathcal{J} =\frac{j}{\sqrt{\lambda}}$, so that $\mathcal{J}\ll 1$ describes short strings while $\mathcal{J}\gg 1$ takes us to long string limit. }


\medskip

\paragraph{Folds on accelerated worldsheet:}
We now come to our case of interest, viz. accelerated strings moving in a locally Rindler spacetime. We will see that the Rindler closed string also folds, but in this case the appearance of folds would point to a very different mechanism since there are no spin degrees of freedom considered here.  

\medskip

Remember, the vanishing of the induced metric (\ref{rindm}) was achieved by taking $\xi\to -\infty$ on the worldsheet. If we consider the accelerated worldsheet is embedded into the Rindler spacetime via some function, this implies the induced metric vanishes when certain points on the worldsheet touches the Rindler horizon at $X'^1  \to -\infty$. Irrespective of the embedding, this clearly marks the points $\s = \pm \t$ as the ones where the metric degenerates, as seen from the map \refb{inverse} \footnote{We again remind the reader that degeneration of the worldsheet metric is a primary property of a string in tensionless phase.}. In analogy with what we discussed above about folded strings, these two points are the ones where ``folds'', now due to acceleration, appear. Now one could think of this as dividing the worldsheet into two causally disconnected sections in the $(\eta,\xi)$ parameterisation corresponding to $\s > \t$ and $\s<\t$ regions. Our previous map \refb{inverse} would be valid only for the $\s>\t$ patch, corresponding to say the right Rindler wedge of the full Minkowski diamond. 

\begin{figure}[t]
\centering
  \includegraphics[width=11cm]{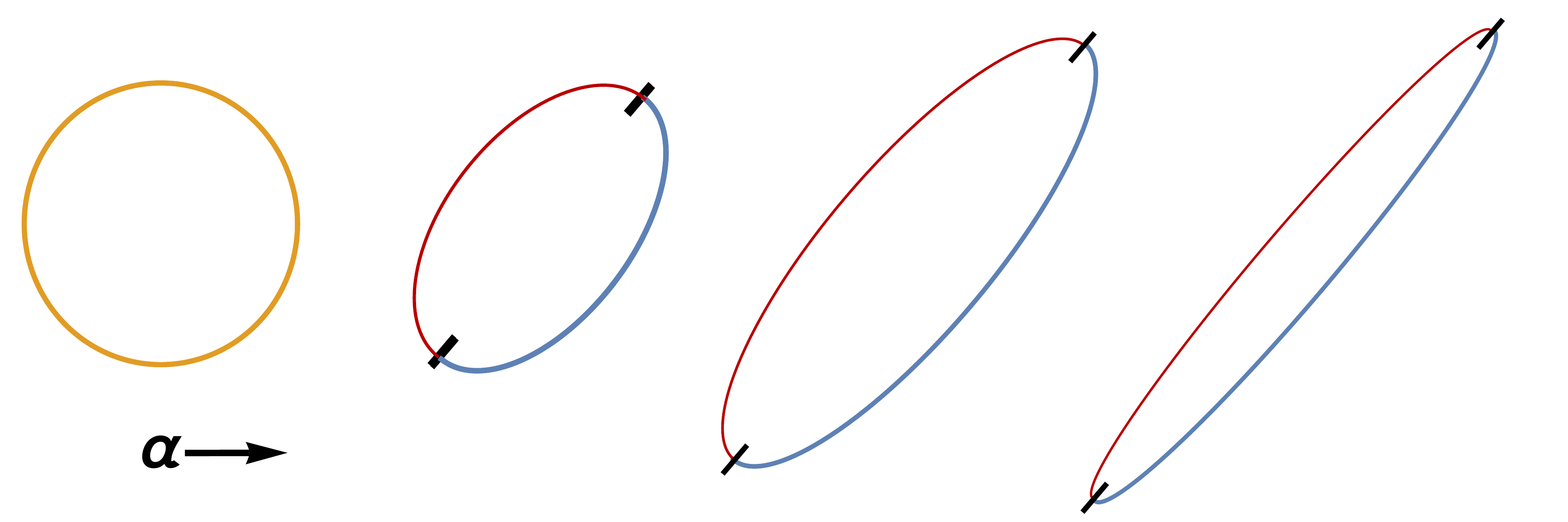}
  \caption{Illustration of how acclereation affects the string shape. As the acceleration increases from zero, the string ``folds'' at the special points and gradually gets more and more elongated. One can think of the special points as boundary of two `open' strings.  }
  \label{}
\end{figure}

\medskip

The accelerated closed string now can be thought of as two open strings ``glued'' together at the folding points with appropriate boundary conditions so as to continue to maintain properties of the closed string like level matching of the states. We will later show these gluing conditions have deeper implications for the physics of such strings.

\subsection{Periodicity}

We now clarify what is going to be a vital point in the rest of our analysis in this paper, viz. the periodicity of the string in the accelerated coordinates $(\eta, \xi)$. The mappings \refb{st} and \refb{inverse} would be central to our discussion. We started with a theory of closed strings, and we want our accelerated string to be still closed after the conformal transformation on the worldsheet. Later in the paper, we will show that this is indeed the case when we look at the evolution of the level matching condition of the string along the flow of acceleration. But for the moment, we focus on the periodicity of the accelerated string  
\footnote{Similar discussions for changing periodicities for strings in Rindler space can be found in \cite{Sanchez:1987yz}, however in a slightly different context.}.
\medskip

\begin{figure}[t]
\centering
  \includegraphics[scale=0.23]{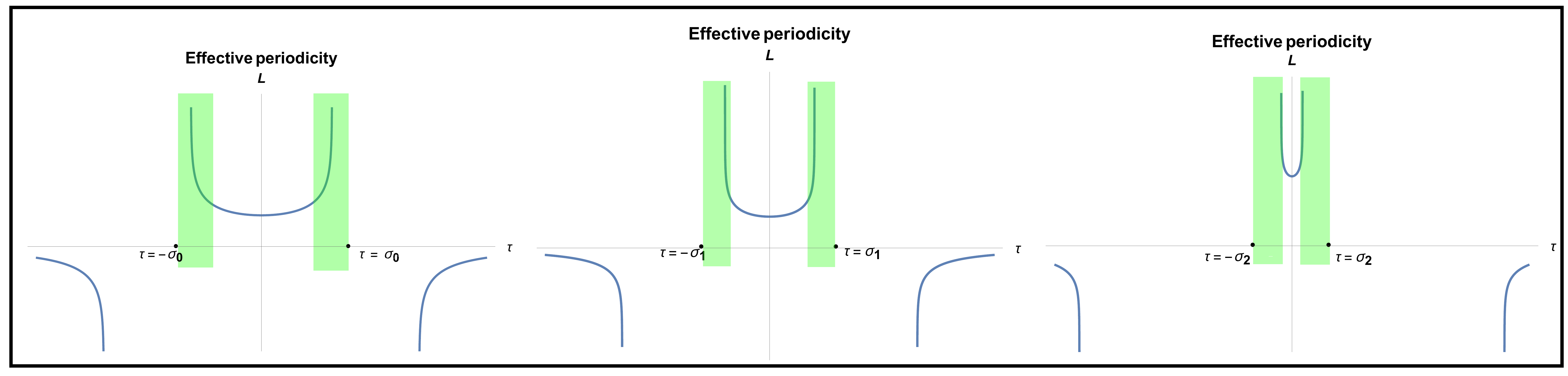}
  \caption{Effective periodicity}
  \label{leff}
\end{figure}

From a target space perspective, the ``closedness" conditions for strings propagating in flat space is given by
\be{}
\Delta\s = 2\pi, ~~\Delta\t = 0,~~X^{\mu}(\s,\tau)=X^{\mu}(2\pi+\s,\tau).
\ee 
With the change in embedding function, these conditions would also change. Since we want closed strings to remain closed in the new embedding, we demand the following:
\bea{}
X^\mu(\s,\t)&=&e^{a X'^\mu(\xi(\s,\tau),\eta(\s,\tau))}~g(a X'^\mu(\xi(\s,\t),\eta(\s,\t)))\\ 
\rightarrow X^\mu(2\pi+\s,\t)&=&e^{a X'^\mu(\xi(2\pi+\s,\tau),\eta(2\pi+\s,\tau))}~g(a X'^\mu(\xi(2\pi+\s,\t),\eta(2\pi+\s,\t))), \nonumber
\eea 
where $g$ is the hyperbolic (co)sine function depending on the embedding coordinates.  Also we have to consider that on the Rindler worldsheet, the spatial periodicity condition of $2\pi$ from the target space embedding will change to an effective periodicity $L_{\text{eff}}$, with,
\begin{align}
    X'^{\mu}(\xi,\eta)&=X'^{\mu}(L_{\text{eff}}+\xi,\eta).
\end{align}
Now we come to a rather subtle point regarding this effective periodicity. Consider the Rindler worldsheet foliated by constant time slices. These can either be inertial time slices of constant $\t$ or Rindler time slices of constant $\eta$. The slicings are evidently very different and only coincide at $\t=\eta=0$ {\footnote{For a visualisation of this, we refer the reader back to Figure~(\ref{rsp}). One can thing of this as a longitudinal cross-section of an accelerating string worldsheet. The constant $\eta$ slices are depicted in the figure. The constant $\tau$ slices can be envisioned as being parallel to the $X$ axis of the figure.}}. So, in order to compare periodicities, we will need to focus on this particular slice. The closedness condition along the spatial or `angular' direction $\xi$ requires
\bea{period}
L_{\text{eff}} = \xi(2\pi,0)-\xi(0,0)&=& \lim_{\delta\to 0}\frac{1}{2a}\left(\log((\delta+2\pi)^2)-\log(\delta^2)    \right)\nonumber \\ 
&=& \lim_{\delta\to 0}~ \frac{1}{a}\log\left(\frac{2\pi}{\delta}+1  \right).
\eea
Here $\delta$ is a parameter which has been introduced to regulate the singularity from the log term, and generically provides a initial value for $\s$. The singularity is an artefact of the horizon which is at $\s=\pm\t$. Since we are looking at $\t=0$, this singularity would hit us in this calculation. Had  we not taken the $\t=0$ slice, the singularity would still appear at $\s=\pm\t$. We illustrate this in Figure (\ref{leff}). 

\medskip

\medskip

The figure shows the effective periodicity at different snapshots of $\s$. The green bands are where the periodicity blows up due to the proximity to the Rindler horizon. These are the regions we will be interested in. The take home message from this toy calculation is the accelerated worldsheet departs from $2\pi$ very fast in terms of spatial periodicity in the proximity of worldsheet horizons. 

\medskip

To account for the regularization we discussed above, one can propose a \textit{regulated} conformal transformation on the worldsheet of the form,
\begin{align}
    \sigma + \delta =e^{a(\xi-\chi)}\cosh{a\eta},~~~ \tau=e^{a(\xi-\chi)}\sinh{a\eta}
\end{align}
where $\chi$ is a corresponding regularization parameter in the Rindler worldsheet. 
If we look at the two coordinate patches $(\s,\t)$ and $(\eta,\xi)$, there is a mismatch between the origins. This can be reconciled provided $\delta = e^{-a\chi}$. This does not hamper the consistency with string equations of motion and constraints as these parameters are simply additive constants to the worldsheet embeddings. Note that for this choice one would get
\be{}
L_{\text{eff}} \approx \frac{1}{a} \log(2\pi)+\chi.
\ee
The inverse transformations in this case reads
\begin{align} \label{inversereg}
\xi-\chi=\frac{1}{2a}\log{[(\sigma+\delta)^2-\tau^2]},~~~ \eta=\frac{1}{2a}\log{\Big[\frac{\sigma+\delta+\tau}{\sigma+\delta-\tau}\Big]}.
   \end{align}
It is easy to see here that as acceleration gets larger and larger, $\delta\to 0$ provided $\chi$ is positive. 

\medskip

The diverging worldsheet periodicity is thus indicative of the presence of horizons on the worldsheet. What we have found in this section is thus concrete signature that the worldheet horizon, which in turn is induced by spacetime horizons, cause changes in the nature of the worldsheet as one moves from inertial worldsheets to accelerated ones. The string still remains closed, but there are marked points appearing on the worldsheet indicating that the initial $S^1$ topology is modified to include defects. These are the folding points we discussed at the beginning of the section. 
\medskip

We make sure that non-inertial periodicities are still finite, but there are singularities at the folding points which are indicative of the closed string becoming two open string segments joined at the folding points when the string is accelerated. However, as mentioned earlier, the memory that the string is still `glued' closed resides even at the quantum level via retention of the level matching conditions, which we discuss later in Sec \ref{sec7}. The boundary conditions at the folding points thus need to be matched properly. As we have been saying for last couple sections, at the very end, when the string becomes null, all points become folding points and we are thus left with an open string.  

\newpage

\section{Worldsheet symmetries and appearance of BMS} \label{sec6}

We now turn our attention to the symmetries of the worldsheet and see what happens to them as we approach the horizon on the Rindler worldsheet. Our view will be predominantly worldsheet symmetry algebra centric in this section and the following discussion will be valid as long as the maps between inertial and accelerated ones are valid. 

\subsection{Accelerated coordinates and Virasoro}
We begin with the inertial string, where the computation of residual gauge symmetry is well known. On a closed string worldsheet, the Virasoro generators are given by ($\tau$, $\sigma$ are world-sheet coordinates)
\begin{align}
    \mathcal{L}_{n}=\frac{i}{2}e^{in(\tau+\sigma)}(\partial_{\tau}+\partial_{\s}),~~~\bar{\mathcal{L}}_{n}=\frac{i}{2}e^{in(\tau-\sigma)}(\partial_{\tau}-\partial_{\s}).
\end{align}
These generators satisfy the classical part of the Virasoro algebra
\begin{align}
     [\mathcal{L}_{m},\mathcal{L}_{n}]=(m-n)\mathcal{L}_{m+n},~~~ [\bar{\mathcal{L}}_{m},\bar{\mathcal{L}}_{n}]=(m-n)\bar{\mathcal{L}}_{m+n},
\end{align}
and one finds the residual symmetry to be two copies of Virasoro. For the inertial closed string, the zero modes of these generators count the number of excitations in the left/right chiral half, and are related via the level-matching condition.  

\medskip

It has been shown in the last section that if the target spacetime is taken to be Rindler (i.e. the string is viewed by a Rindler observer with constant acceleration $a$) then at the level of worldsheet coordinates we can have the following transformation
\begin{align}
    \sigma\pm\tau+\delta=\pm e^{a(\xi\pm\eta-\chi)},
\end{align}
where the parameters $\delta$ and $\chi$ are introduced in order to ensure a finite string period. We will ignore these parameters for the purposes of this calculation (their inclusion do not alter results). The $+/-$ sign corresponds to transformations for Right/Left wedges, which are now causally disconnected sectors from each other.
The Virasoro generators for the above reparametrization in the Right wedge become
\begin{subequations}\label{L1}
\begin{align}
     \mathcal{L}_{n}^{(R)}&=\frac{i}{2a}\exp{[ine^{a(\xi+\eta)}-a(\xi+\eta)]}(\partial_{\eta}+\partial_{\xi}) \\
     \bar{\mathcal{L}}_{n}^{(R)}&=\frac{i}{2a}\exp{[-ine^{a(\xi-\eta)}-a(\xi-\eta)]}(\partial_{\eta}-\partial_{\xi}).
\end{align}
\end{subequations}
Similarly in the Left wedge the generators take the following form
\begin{subequations}\label{L2}
\begin{align}
     \mathcal{L}_{n}^{(L)}&=-\frac{i}{2a}\exp{[-ine^{a(\xi+\eta)}-a(\xi+\eta)]}(\partial_{\eta}+\partial_{\xi}) \\
     \bar{\mathcal{L}}_{n}^{(L)}&=-\frac{i}{2a}\exp{[ine^{a(\xi-\eta)}-a(\xi-\eta)]}(\partial_{\eta}-\partial_{\xi}).
\end{align}
\end{subequations}
Both of these sets of generators at finite acceleration again satisfy the Virasoro algebra as in the case of the inertial string.   We highlight an important point here. The generators in the two wedges are related via a``Flipping" i.e.  
\be{flip}
\mathcal{L}_{n}^{(R)}\to -\mathcal{L}_{-n}^{(L)}, \quad \mathcal{L}_{n}^{(L)}\to -\mathcal{L}_{-n}^{(R)}
\ee
and similarly for the anti-holomorphic one. This happens because the timelike killing vector changes direction when one moves from one wedge to another. Since this transformation is an automorphism of the Virasoro algebra, the worldsheet symmetry structure remains invariant in finite acceleration. 

\subsection{Contractions and BMS} 
We will now work with units where the acceleration is constant, or without losing generality, $a=1$ \footnote{This is very similar to the way one takes e.g. the non-relativistic limit by scaling coordinates $t\to t$ and $x^i\to \e x^i$ with $\e\to0$ while setting the speed of light $c=1$ \cite{Bagchi:2009my}.}. We are interested in approaching the horizon in the 2d Rindler worldsheet. Remembering that trajectories at constant $\xi$ are also trajectories at constant acceleration, as we approach the horizon, $\xi\to -\infty$ or equivalently the acceleration goes to infinity. To see this algebraically, we will perform the contraction 
\be{etacont}
\xi\to\xi \quad \text{and} \quad \eta\to\epsilon\eta  \quad \text{with} \quad \e\to0.
\ee 
This effectively makes $\xi$ very large. In the limit $\epsilon\to 0$ we can consider the following contracted generators in the Right wedge
\begin{subequations}\label{genbms}
\begin{align}
    L_{n}&=\mathcal{L}_{n}^{(R)}-\bar{\mathcal{L}}_{-n}^{(R)}=i\exp{(ine^\xi-\xi)}[\partial_{\xi}+(ine^{\xi}-1)\eta\partial_{\eta}] \\
    M_{n}&=\epsilon(\mathcal{L}_{n}^{(R)}+\bar{\mathcal{L}}_{-n}^{(R)})=i\exp{(ine^\xi-\xi)}\partial_{\eta}.
\end{align}
\end{subequations}
These contracted generators $L_{n}$ and $M_{n}$ can be shown to satisfy the classical part of BMS$_3$ algebra given by
\begin{align}
    [L_{m},L_{n}]=(m-n)L_{m+n},~~[L_{m},M_{n}]=(m-n)M_{m+n},~~[M_{m},M_{n}]=0.
\end{align}
The contraction \refb{etacont} can also be understood from the mapping between the inertial and accelerated coordinates \refb{inverse} and putting \refb{URlim} in  there. Similar contractions in the Left wedge gives analogous generators with the flipping transformations (\ref{flip}) in place, i.e. $L_{n}\to -L_{-n}$ and $M_{n}\to -M_{-n}$. Sitting on either wedge, these limits appear to be the Ultrarelativistic (UR) or Carrollian as discussed before (\ref{vir2bms}). This is further evident when we remember that $\eta$ is the Rindler time and hence even in this case \refb{etacont} contracts the time direction and hence is Carrollian.

\begin{figure}[h]
\centering
  \includegraphics[width=14.5cm]{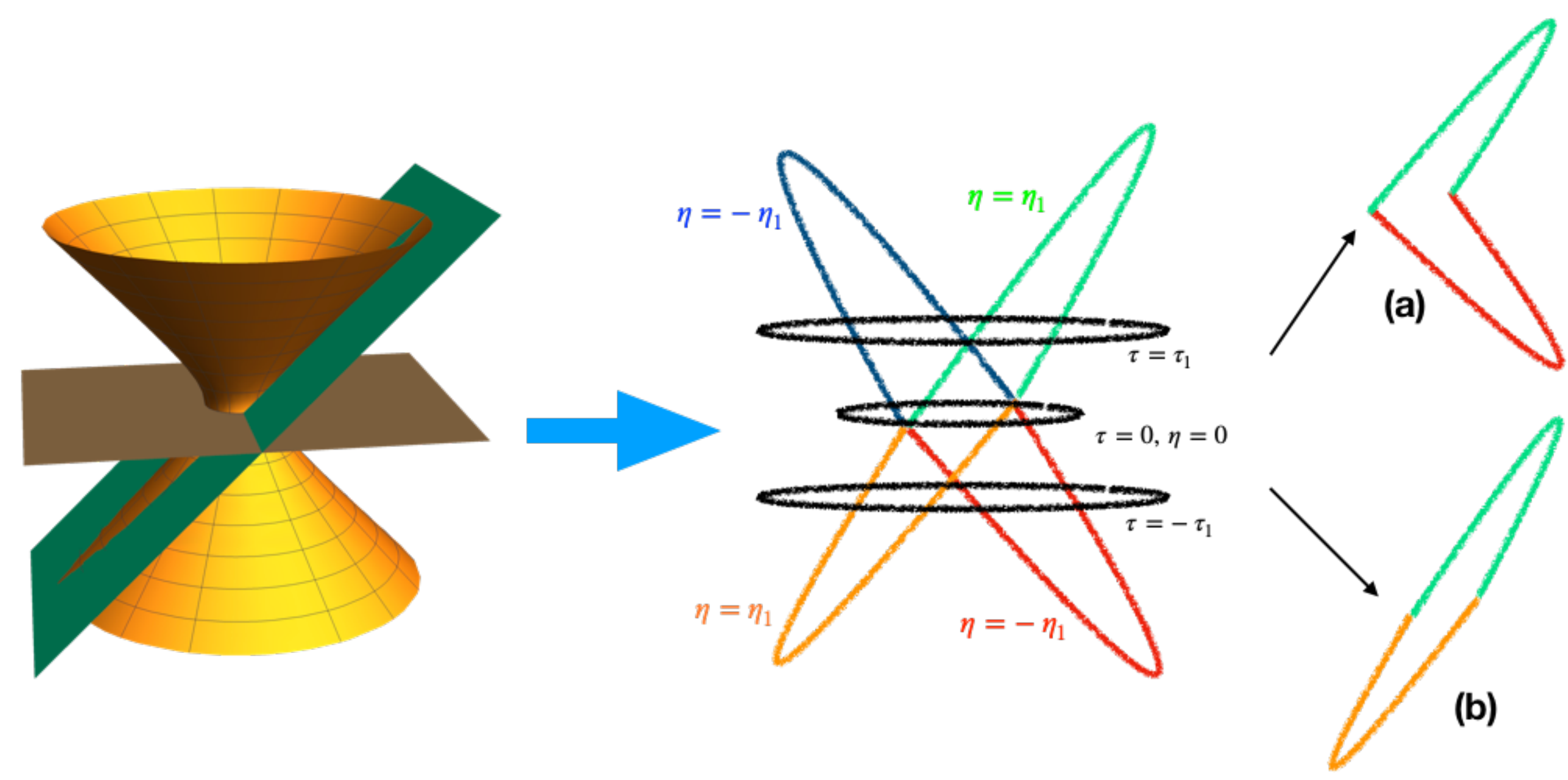}
  \caption{An idealised illustration of constant timeslices in inertial and non-inertial case. Notice that the inertial $\t$ and non-inertial $\eta$ both coincide only at the initial slice and nowhere else. The three black sections correspond to inertial slices at $0,\pm\t_1$, while the green and orange (blue and red) segments make up the $\eta_1$  ($-\eta_1$) non-inertial elliptical slices. Inset we show ways to recombine the segments into `closed' strings. }
  \label{segment}
\end{figure}

\medskip

Let us now move to a rather intriguing observation. We can achieve this contraction in another way by using the map \refb{flip}. In this case, we will combine the holomorphic generators of the right wedge with the anti-holomorphic generators on the left wedge. 
\begin{align}\label{newcontract}
    L_{n}=\mathcal{L}_{n}^{(R)}+\bar{\mathcal{L}}_{n}^{(L)},~~~M_{n}=\epsilon(\mathcal{L}_{n}^{(R)}-\bar{\mathcal{L}}_{n}^{(L)}).
\end{align}
Here clearly the ``right'' and ``left'' modes mix together and create the new set of generators the form of which are exactly identical to \refb{genbms} due to the isomorphism indicated above. This particular contraction looks like a Non-relativistic (NR) one \refb{NR}. This BMS algebra has legs in the both wedges of the accelerated worldsheet, instead of the algebra defined on the single wedge as we defined before. We now try to explain this new and intriguing observation.

\medskip

To understand this let us look at the figure (\ref{segment}). This depicts the worldsheet of an accelerated string, which can be represented by a hyperboloid. Here we look at slicing this hyperboloid, first in constant inertial times $\t$ and then in Rindler times $\eta$. The inertial slices are the horizontal slices where the cross-sections are all circular and smoothly connected. The Rindler slices on the other hand give us ellipses of higher and higher eccentricity, ultimately reaching an infinite line when the Rindler horizon is hit. 

\medskip

We now look at the Rindler slices at two instances $\eta=\pm \eta_1$. These are given by a green and orange closed string (for $\eta=\eta_1$) and a blue and red closed string (for $\eta=-\eta_1$). The four segments of different colours are joined at two (`folding') points at the $\eta=\t=0$ slice. The important thing to notice here is the Rindler map \refb{inverse} does not simultaneously encompass all the segments in Figure (\ref{segment}) as $\t \to -\t$ implies $\eta \to -\eta $ in there. Specifically, in inset (a) the green and red segments are covered by this map, and similarly for the blue and orange segments, or in other words, segments only in the R wedge or in L wedge. So in order to form a closed string in either of the Rindler wedges, we need to take the actual closed string (the green-orange one in the figure, inset (b), encompassing both wedges) and take a transformation of the lower half (the orange segment) from the left wedge to the right wedge, thus transforming this to the red segment. The Right Rindler closed string is thus the green-red string depicted in part (a) of Figure (\ref{segment}). 

\medskip

Our first contraction \refb{genbms} is the process by which this green-red string becomes null as we approach the Rindler horizon from the right wedge. This is the evolution in acceleration of an ensemble of further and further accelerated Rindler strings. Our second contraction \refb{newcontract} is the Rindler time evolution of a closed (green-orange) string as it gets closer and closer to the horizon. This encompasses both wedges and is connected to defining global modes which we will describe in the coming section. 

\newpage

\section{Quantum states on the accelerated worldsheet}\label{sec7}
In this section, we discuss how to define vacua and excited states on the accelerated worldsheet. We will describe how to write out the mode expansions for this string and how the actions of the oscillators are defined in this setting.

\subsection{Mode expansions and Bogoliubov transformations}

In case of tensile closed strings with worldsheet coordinates $(\t,\s)$, the worldsheet equation of motion reads 
$$(-\p_{\t}^2+\p_{\s}^2)X^{\mu}=0.$$ 
So, in the inertial frame one can write down the mode expansion in the following form
\begin{equation}
    X^{\mu}(\sigma,\tau)=x^{\mu}+ \alpha' p^{\mu}\tau+\sqrt{2\pi\alpha'}\sum_{n> 0}[\alpha^{\mu}_{n}u_{n}+\alpha^{\mu}_{-n}u^{*}_{n}+\Tilde{\alpha}^{\mu}_{n}\Tilde{u}_{n}
    +\Tilde{\alpha}^{\mu}_{-n}\Tilde{u}^{*}_{n}]
    \label{minkws}
\end{equation}
where the right and left moving mode functions are given by exponentials
\be{smallu}
u_{n}=\frac{ie^{-in(\tau-\sigma)}}{\sqrt{4\pi}n},~~ \Tilde{u}_{n}=\frac{ie^{-in(\tau+\sigma)}}{\sqrt{4\pi}n}.
\ee 
Also, $x^\mu$ and $p^\mu$ correspond to zero modes for the string. Hermitian conjugates of the oscillators are given by $\a^\dagger_{n} = \a_{-n}$ and similarly for the anti-holomorphic one. These oscillators annihilate the inertial closed string vacuum:
\be{alpha1}
\a_n|0\rangle_{\a}= \tilde\a_n|0\rangle_{\a}=0.
\ee

Similarly, in a uniformly accelerating frame, closed string equation of motion assumes the form 
$$(-\p_{\eta}^2+\p_{\xi}^2)X'^{\mu}=0$$ 
due to the conformal transformation (\ref{st}) between the two coordinate systems. Depending on the wedge, we seemingly have two different possible sets of oscillators, $(\beta^{(R)},\tilde\beta^{(R)})$  and $(\beta^{(L)},\tilde\beta^{(L)})$ to make up for two different set of Virasoro generators as discussed in the preceeding section. Keeping ourselves restricted, one could envision the mode expansions in terms of only Right wedge oscillators,
\bea{}
X(\xi, \eta) &=& X_{CM}+ \sqrt{2 \pi\alpha'}\sum_{n>0}[\b_n U_n+\tilde{\b}_n \tilde{U}_n+h.c.].
\eea
 But as we discussed in the last section, there is a clear way to have oscillators from both wedges to contribute in the residual symmetry algebra (\ref{newcontract}), which must be reflected in the mode expansions as well,  thereby making the system a `closed' string. In this sense, the mode expansion valid over the whole Rindler worldsheet \footnote{Referring to figure (\ref{segment}), this will be the green-orange string lying on a constant $\eta$ timeslice in inset (b). } is written schematically as
\begin{equation} 
    X'^{\mu}(\xi,\eta)=q^{\mu}+\alpha'p^{\mu}\eta+\sqrt{2\pi\alpha'} \sum_{n> 0}[\beta^{\mu(R)}_{n}U_{n}^{(R)}+\tilde\beta^{\mu(L)}_{n}{U}_{n}^{(L)}+h.c.].
\label{rindlerws}
\end{equation}
where the mode functions are defined as
\begin{eqnarray}
U_{n}^{(R)}=\frac{ie^{-i\lambda_{n}(\xi-\eta)}}{\sqrt{4\pi}n}, ~~{U}_{n}^{(L)}=\frac{ie^{-i\lambda_{n}(\xi+\eta)}}{\sqrt{4\pi}n}.
\end{eqnarray}
These are standard Rindler space mode functions, where the frequencies are positive with respect to the flow of time in the two wedges. Bear in mind $U_{n}^{(R)}$ is only defined in the Right wedge and $U_{n}^{(L)}$ is only defined in the Left one. Also $q^\mu$ and $p^\mu$ are usual center of mass coordinates and momenta\footnote{Since we have two causally disconnected segments, one might want to use different center of mass coordinates for the L and R segment. We will come back to this later.}. 

\medskip

Also note that here, the frequency of the mode functions and the periodicity are modified as compared with the ones given in (\ref{smallu}) and take the respective forms $\lambda_{n}=\frac{2\pi n}{L_{\delta}}$ and $X'^{\mu}(0,\eta)=X'^{\mu}(L_{\delta},\eta)$.  Here $L_{\delta}$ is the changed string periodicity due to the `defect' as we had discussed in (\ref{period}). It is to be understood that we are dealing with the parameter space corresponding to the green bands in figure (\ref{leff}).

\medskip

All the sets of oscillators individually have to follow canonical commutation relations i.e. 
$$[\b_n,\b_m]=n\delta_{m+n}=[\tilde\b_n,\tilde\b_m]$$ 
and similarly for the other sets. This covers both the wedges of accelerated worldsheet in consideration. The Virasoro generators are simply the bilinears of the individual oscillators which can be computed using constraints from the classical action. In this case they look like the following:
\be{}
\mathcal{L}_{n}^{(R)}= \frac{1}{2}\sum_{m}\b_{-m}^{(R)}\cdot \b_{m+n}^{(R)}~~~\mathcal{\bL}_{n}^{(L)}= \frac{1}{2}\sum_{m}\tilde\b_{-m}^{(L)}\cdot \tilde\b_{m+n}^{(L)}.
\ee
We will have similar definitions for the (anti)holomorphic ones. Now since we found in the last section that generators in the string segments of two wedges transform into each other via a flipping isomorphism, which ensures $\mathcal{\bL}_{n}^{(L)}\to -\mathcal{\bL}_{-n}^{(R)}$, the anti-holomorphic set of oscillators then can be related across wedges via the following analytical continuations: 
\be{}
\tilde\b_{n}^{(L)}\to i\tilde\b_{-n}^{(R)}.
\ee

\begin{figure}[h]
\centering
  \includegraphics[width=8.5cm]{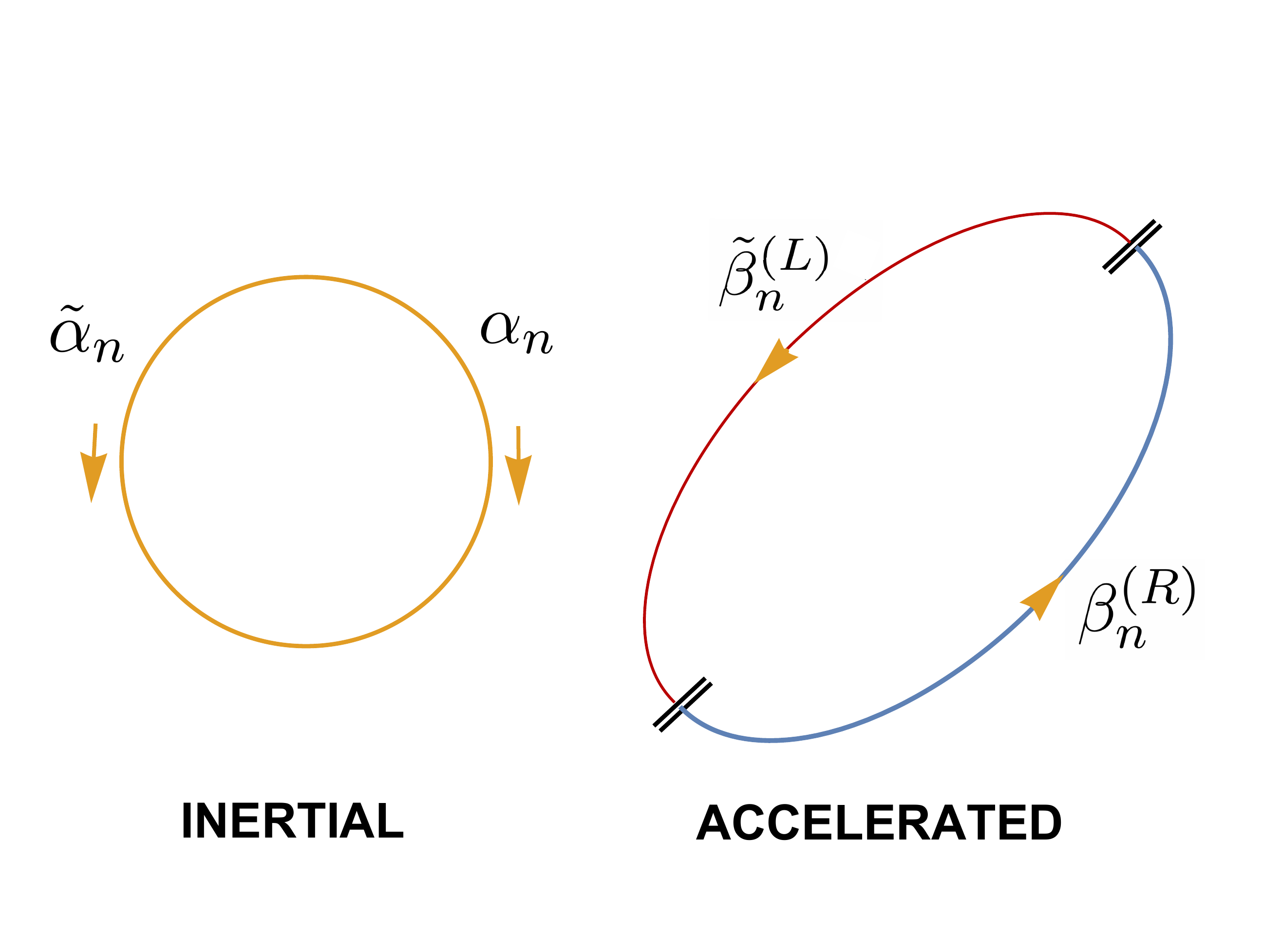}
  \caption{Illustration of temporal cross sections for inertial and accelerated worldsheets. The arrows indicate flow of timelike killing fields. Accelerated string is ``divided'' into two segments due to worldsheet discontinuities, but continues to ``live'' simultaneously in both wedges. }
  \label{}
\end{figure}

So in terms of action on a string state, the creation and annihilation operators change into one another as we cross from one wedge to the other. 
As one can verify, the above transformations keep the canonical commutation relations intact. This also summarizes the mechanism of UR $\leftrightarrow$ NR duality we discussed in relation with the possible contractions  \refb{genbms}, \refb{newcontract}. 
 It goes without saying that $\beta$ oscillators annihilate a different vacuum from the inertial one,  which encompasses both wedges as a direct product:
\be{}
\b_n^{(R)}|0\rangle_{\b}= \tilde\b_n^{(L)}|0\rangle_{\b}=0.~~~~|0\rangle_{\b} \equiv |0\rangle_{R}\otimes |0\rangle_{L}.
\ee

The reader is here reminded that in (\ref{smallu}), $u_{n}$ and $\tilde{u}_{n}$ are positive energy mode functions defined everywhere on the worldsheet of a closed string described by the coordinates ($\tau, \sigma$). On the other hand, $U_{n}^{(R)}$ and $U_{n}^{(L)}$ are positive energy mode functions defined in two causally disconnected regions of the closed string worldsheet respectively. These two causally disconnected regions can be described as left and right wedges in the following way, 
\begin{align}
 &&    \sigma+\delta = e^{a(\xi-\chi)}\cosh{a\eta},~ \tau = e^{a(\xi-\chi)}\sinh{a\eta} ~~~ \text{(\textit{Right} wedge)}\nonumber \\
 &&       \sigma+\delta = -e^{a(\xi-\chi)}\cosh{a\eta}, ~\tau = -e^{a(\xi-\chi)}\sinh{a\eta} ~~~ \text{(\textit{Left} wedge)} 
\end{align}
where ($\eta$, $\xi$) are the local worldsheet coordinates in each wedge. 

\medskip

\begin{figure}[h]
\centering
  \includegraphics[width=11.5cm]{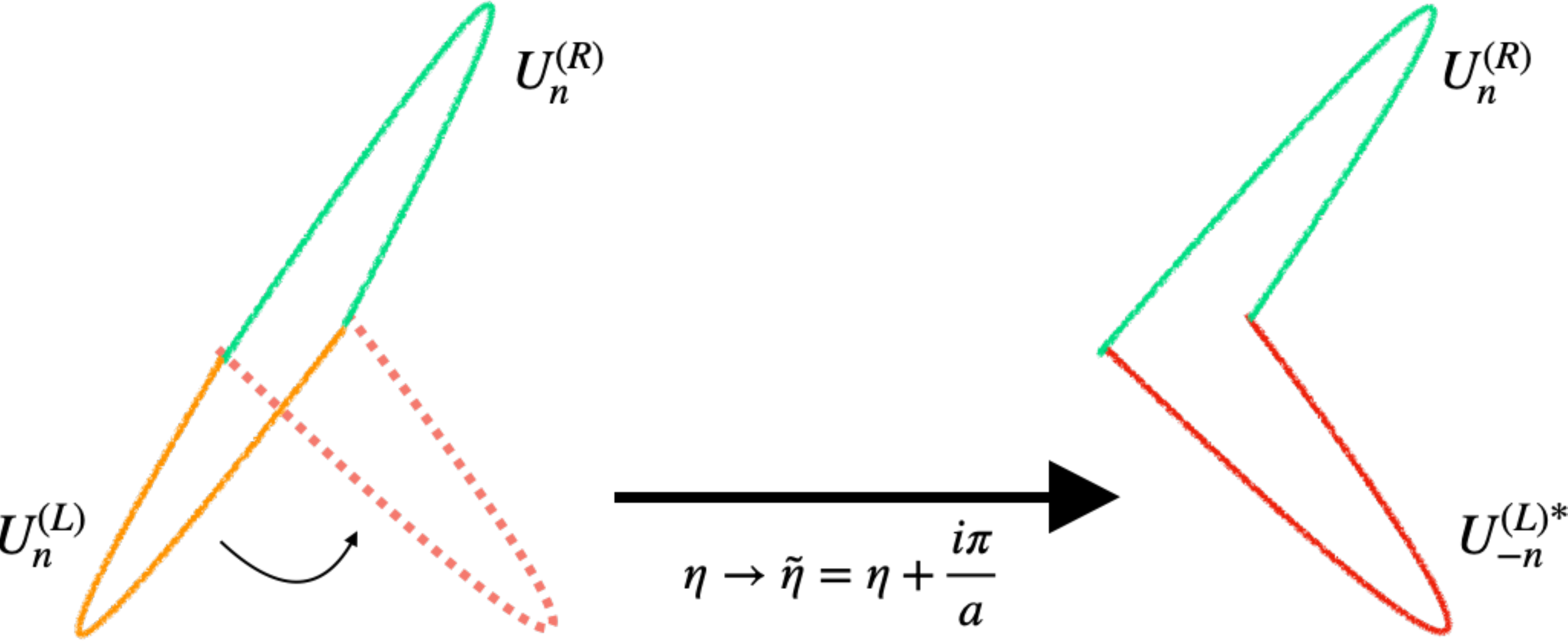}
  \caption{A visual representation of defining global modes of the accelerated string. Here the green and orange segments are on a constant $\eta$ slice, hence an actual `closed' string. An analytical continuation on Rindler time maps the orange segment into the red one, projecting it into the Right Rindler string. This combination of modes are globally well-defined.  }
  \label{segway}
\end{figure}

Keeping in line with scalar field quantization in Rindler space \cite{Birrell:1982ix}, we now need to analytically extend $U_n^{(R)}$ and $U_n^{(L)}$ to find a global positive frequency mode defined at both the wedges. In order to do so, let us re-express the local mode of right wedge in terms of global worldsheet coordinates ($\tau$,$\sigma$),
\be{RU} 
    \sqrt{4\pi}n \ U_{n}^{(R)} =ie^{-i\lambda_{n}(\xi-\eta)}=i(\sigma-\tau+\delta)^{\frac{-i\lambda_{n}}{a}}e^{-i\lambda_{n}\chi}
\ee
Similarly in the left wedge we have 
\be{LU}
     \sqrt{4\pi}n \ U^{(L)*}_{-n}=ie^{-i\lambda_{n}(\xi-\eta)}=i(-1)^{\frac{-i\lambda_{n}}{a}}(\sigma-\tau+\delta)^{\frac{-i\lambda_{n}}{a}}e^{-i\lambda_{n}\chi}.
\ee
Therefore the following combination 
\begin{equation}\label{globmode}
   U_{n}^{(R)}-e^{-\frac{\pi\lambda_{n}}{a}}U^{(L)*}_{-n}= i(\sigma-\tau+\delta)^{\frac{-i\lambda_{n}}{a}}e^{-i\lambda_{n}\chi}
\end{equation}
defines a global mode that takes care of both left and right wedges, and hence can be considered as analytic extension of $U_{n}^{(L)}$ to the Right wedge. Looking at the analytic formulas, one can easily spot that the two terms in the LHS of (\ref{globmode}) are related to each other by an analytical continuation of the Rindler time  
\be{}
\eta \to \tilde\eta = \eta+\frac{i\pi}{a}.
\ee
Rather intriguingly, looking at the symmetry generators in (\ref{L1})-(\ref{L2}), we can see this temporal analytical continuation maps $\bL_{n}^{(L)}$ into $\bL_{n}^{(R)}$. So in a way, the above definition of the Right Wedge global mode actually maps the cumulative physics into the Right Rindler string, which we have discussed about before. To show this clearly, let us borrow elements from figure (\ref{segment}) and point this transformation out in  figure (\ref{segway}). The reader is reminded that this transformation is different from the flipping transformations we mentioned before simply by definition. 

Similarly as above, we can find the analytic extension of $U_{n}^{(R)}$ into the Left wedge as well which turns out to have a similar form
\begin{equation}
U_{n}^{(L)}-e^{-\frac{\pi\lambda_{n}}{a}}U^{(R)*}_{-n}.
\end{equation}
This represents a mirror-reflected configuration of the green-red string in figure (\ref{segway}) into the Left wedge. After normalizing, we now get a set of global modes with support in both wedges \footnote{The reader is directed to Appendix \ref{ApA} for details on these modes.}: 
\begin{align}
h_{n}&=\frac{1}{\sqrt{2\sinh{\Big(\frac{\pi\lambda_{n}}{2a}\Big)}}}\Big(e^{\frac{\pi\lambda_{n}}{2a}}U_{n}^{(R)}-e^{-\frac{\pi\lambda_{n}}{2a}}{U}^{(R)*}_{-n}\Big)\nonumber \\
\Tilde{h}_{n}&=\frac{1}{\sqrt{2\sinh{\Big(\frac{\pi\lambda_{n}}{2a}\Big)}}}\Big(e^{\frac{\pi\lambda_{n}}{2a}}{U}^{(L)}_{n}-e^{-\frac{\pi\lambda_{n}}{2a}}U^{(R)*}_{-n}\Big)
\label{normglob}
\end{align}
Since the mode functions given above are well-defined in entire closed string worldsheet, we can choose them as an independent basis in the expansion of embedding function $X^{\mu}$.  
 \begin{equation}
    X^{\mu}(\xi,\eta)=q_{RG}^{\mu}+\alpha'p_{RG}^{\mu}\eta+\sqrt{2\pi\alpha'} \sum_{n> 0}[\gamma^{\mu}_{n}h_{n}+\gamma^{\mu}_{-n}h^{*}_{n}+\Tilde{\gamma}^{\mu}_{n}\Tilde{h}_{n}+\Tilde{\gamma}^{\mu}_{-n}\Tilde{h}^{*}_{n}],
\label{globalws}
\end{equation}
 Note that although the string coordinate in (\ref{globalws}) is defined as an embedding function of local coordinate ($\eta$, $\xi$), the mode expansion is valid in both wedges unlike the one presented before. Also, ($h_n, \Tilde{h}_n$) satisfy the same positive frequency analyticity property as ($u_n$, $\tilde{u}_n$) and they share a common vacuum state as argued in \cite{Unruh:1976db}. ($\gamma_n$, $\tilde{\gamma}_n$) are global oscillators which still annihilate the inertial string vacua,
 \be{}
\gamma_n|0\rangle_{\a}= \tilde\gamma_n|0\rangle_{\a}=0.
\ee
 Generically then $\gamma$ oscillators will be a linear combination of the $\a$ oscillators that also define the inertial closed string vacuum (\ref{alpha1}). For all purposes, we can choose a basis where they coincide sans some constant diagonalizing factor. Using the relations (\ref{normglob}) in (\ref{rindlerws}), and comparing it with (\ref{globalws}) as both of them stand for the expansion of same embedding function, we arrive at the Bogoliubov transformation between the accelerated oscillators ($\beta_n$, $\tilde{\beta}_n$) and the inertial oscillators ($\a_n$, $\tilde{\a}_n$)
\begin{subequations}  \label{bv}
\begin{align}
    \beta^{\mu(R)}_{n}&=\frac{e^{\frac{\pi\lambda_{n}}{2a}}}{\sqrt{2\sinh{\Big(\frac{\pi\lambda_{n}}{2a}\Big)}}}\a^{\mu}_{n}-\frac{e^{-\frac{\pi\lambda_{n}}{2a}}}{\sqrt{2\sinh{\Big(\frac{\pi\lambda_{n}}{2a}\Big)}}}\Tilde{\a}^{\mu}_{-n} \\
    \Tilde{\beta}^{\mu(L)}_{n}&=-\frac{e^{-\frac{\pi\lambda_{n}}{2a}}}{\sqrt{2\sinh{\Big(\frac{\pi\lambda_{n}}{2a}\Big)}}}\a^{\mu}_{-n}+\frac{e^{\frac{\pi\lambda_{n}}{2a}}}{\sqrt{2\sinh{\Big(\frac{\pi\lambda_{n}}{2a}\Big)}}}\Tilde{\a}^{\mu}_{n}
 \end{align}
\end{subequations}
These are self-consistent in the sense canonical commutation relations don't change under these transformations.
Keeping in mind that $\lambda_{n}=\frac{2\pi n}{L_{\delta}}$ and $L_{\text{eff}}=\frac{1}{a}\log\Big[\frac{2\pi}{\delta}+1\Big]$ we see that 
\begin{equation}
    \frac{\pi \lambda_{n}}{2a}=\frac{\pi^2 n}{\log \Big[\frac{2\pi}{\delta}+1\Big]}.
    \label{lambdel}
\end{equation}
For $\delta\to 0$ we see that the denominator of r.h.s. in (\ref{lambdel}) diverges logarithmically and hence the combination $ \frac{\pi \lambda_{n}}{2a}$ approaches to zero. In the limit of small $ \frac{\pi \lambda_{n}}{2a}$, the Bogoliubov coefficients in (\ref{bv}) admit the following form,
\begin{subequations}\label{bvsimplified}
\begin{align} 
\beta^{\mu(R)}_{n} &= \frac{1}{2}\Bigg[\sqrt{\frac{\pi^2 n}{\log \Big[\frac{2\pi}{\delta}+1\Big]}}+\sqrt{\frac{1}{\pi^2 n}\log \Big[\frac{2\pi}{\delta}+1\Big]}\Bigg]\a^{\mu}_{n} \nonumber \\ &\hspace{4 cm}+\frac{1}{2}\Bigg[\sqrt{\frac{\pi^2 n}{\log \Big[\frac{2\pi}{\delta}+1\Big]}}-\sqrt{\frac{1}{\pi^2 n}\log \Big[\frac{2\pi}{\delta}+1\Big]}\Bigg]\Tilde{\a}^{\mu}_{-n} \\
 \Tilde{\beta}^{\mu(L)}_{n} &= \frac{1}{2}\Bigg[\sqrt{\frac{\pi^2 n}{\log \Big[\frac{2\pi}{\delta}+1\Big]}}-\sqrt{\frac{1}{\pi^2 n}\log \Big[\frac{2\pi}{\delta}+1\Big]}\Bigg]\a^{\mu}_{-n} \nonumber \\ &\hspace{4 cm}+\frac{\pi^2 n}{2}\Bigg[\sqrt{\frac{1}{\log \Big[\frac{2\pi}{\delta}+1\Big]}}+\sqrt{\frac{1}{\pi^2 n}\log \Big[\frac{2\pi}{\delta}+1\Big]}\Bigg]\Tilde{\a}^{\mu}_{n}.
\end{align}
\end{subequations}
Now, recall that $\delta=e^{-a \chi}$ and one of the ways to achieve $\delta\to 0$ is by keeping $\chi>0$ with $\alpha\to\infty$ and as a consequence of that we can further simplify, 
\begin{equation}
\log\Big[\frac{2\pi}{\delta}+1\Big]=\log\Big[\frac{2\pi}{e^{-a \chi}}+1\Big]\to\log e^{a \chi}+\log2\pi\approx a \chi.
\label{delzero}
\end{equation}
Implementing the infinite acceleration limit in (\ref{bvsimplified}) gives,
\begin{subequations}
\begin{align}
    \beta^{\infty(R)}_{n}&=\frac{1}{2}\Bigg[\sqrt{\frac{\pi^2 n}{a \chi}}+\sqrt{\frac{a \chi}{\pi^2 n}}\Bigg]\a_{n}+\frac{1}{2}\Bigg[\sqrt{\frac{\pi^2 n}{a \chi}}-\sqrt{\frac{a \chi}{\pi^2 n}}\Bigg]\Tilde{\a}_{-n} \\
     \Tilde{\beta}^{\infty(L)}_{n}&=\frac{1}{2}\Bigg[\sqrt{\frac{\pi^2 n}{a \chi}}-\sqrt{\frac{a \chi}{\pi^2 n}}\Bigg]\a_{-n}+\frac{1}{2}\Bigg[\sqrt{\frac{\pi^2 n}{a \chi}}+\sqrt{\frac{a \chi}{\pi^2 n}}\Bigg]\Tilde{\a}_{n}.
     \label{infacc}
\end{align}
\end{subequations}


The infinite acceleration limit essentially suggests to zoom in to the case where worldsheet simply hits the lightcone. 
Now looking at the Bogoliubov transformation above, one could easily compare it with the one we got from the theory of tensionless strings in the infinite boost limit (\ref{infvel}). That leads one to infer, upto constant factors, 
\be{reln}
\e \sim \frac{1}{a},
\ee
when one identifies the oscillators in the tensionless regime (see e.g. Eq \refb{infvel}) and maximum acceleration regime \footnote{Zero modes for $C$ have to be treated differently since they are equal. }
\be{}
C_n = \b_n^{\infty(R)},~~\tilde{C}_n =  \Tilde{\beta}^{\infty(L)}_{n}.
\ee
In this case, the tensionless string vacuum $(\e \to 0)$ turns out to be the emergent vacuum for the accelerated string in infinite $a$ limit. For these theories, increasing acceleration amounts to decreasing string tension, and the emergent physics matches at the extreme limits. 

\medskip

To summarize, we have learnt that the accelerated worldsheet oscillators $(\b,\tilde{\b})$ at $a\to \infty$ limit will be comparable to the tensionless oscillators $(C,\C)$ that we discussed in Sec \ref{sec2}. 
\be{}
a\to \infty : \quad \{ \b_n^{(R)}, \tilde{\b}_n^{(L)} \} \to \{C_n, \C_n\}
\ee
We have thus shown that the \textit{physics of highly accelerated strings and that of the tensionless ones are in one to one correspondence}.

\subsection{Closure of accelerated strings}\label{7.2}

We have reemphasised before that despite the discontinuities on the worldsheet the accelerated string still behaves as a closed one. A pointer to this fact is that the residual symmetry algebra on the finitely accelerated worldsheet continues to be two copies of the Virasoro algebra. We now quantify the `closed' property of the accelerated strings in terms of the evolution of level matching conditions and zero modes of the Virasoro generators in accelerated frame.  

\medskip

For the tensile worldsheet, the Hamiltonian and Angular momentum operators are given by combinations of Virasoro zero modes
\be{}
H_M=\L_0+\bL_0,~~~J_M=\L_0-\bL_0
\ee
In the tensile string case, invariance under $J_M$ signifies level matching condition for a certain state. This is a unique feature of closed strings.  Remember, the operators $\L_0$, $\bL_0$ are given in terms of inertial oscillators by 
\be{} \L_0=\frac{1}{2}\sum_m \a_{-m}\cdot\a_{m},~~
\bL_0=\frac{1}{2}\sum_m \ta_{-m}\cdot\ta_{m} \quad. \ee

In the Rindler space, when our observers are accelerated, the oscillators change under Bogoliubov transformations. So the zero mode virasoro generators on an accelerated string will take the following (normal ordered) forms: 
 \begin{subequations}\label{29}
\bea{}
 \L_0^{(\a)}&=\sum_{m} \left[ \Lambda_+^2 \b_{-m}\cdot \b_{m}+\Lambda_-^2 \tilde{\b}_{-m}\cdot \tilde{\b}_{m}-2\Lambda_+\Lambda_- \b_{m}\cdot \tilde{\b}_{m} \right], \\ 
\bar{\L}_0^{(\a)}&=\sum_{m} \left[ \Lambda_+^2 \tilde{\b}_{-m}\cdot \tilde{\b}_{m}+\Lambda_-^2 \b_{-m}\cdot \b_{m}-2\Lambda_+\Lambda_- \b_{m}\cdot \tilde{\b}_{m} \right], \eea
\end{subequations}
where the Bogoliubov transformed oscillators are generically given by 
\bea{30}
\a^{\mu}_n&=\Lambda_+\b^{\mu(R)}_n-\Lambda_-\tilde{\b}^{\mu(L)}_{-n}\nonumber \\
\ta^{\mu}_n&=\Lambda_+\tilde{\b}^{\mu(L)}_{n}-\Lambda_-\b^{\mu(R)}_{-n} ,
 \eea
Eqs \refb{30} are the inverse transforms of (\ref{bv}), $\Lambda_\pm$ are the Bogoliubov coefficients, and we have further dropped the $R/L$ superscript in defining the Virasoro generators. Note that first two oscillator bilinears in  $\L_0^{(\a)}$ and $\bar{\L}_0^{(\a)}$ in \refb{29} are just the number operators acting in the two wedges. It is easy to see that in the limit of infinite acceleration, where we have the identifications $\b_n^\infty =C_n$ and $\tilde{\b}_n^{\infty} = \C_n$, we could use (\ref{reln}), viz. the one-to-one map between $\e$ and $a$,  to go back to the definitions in terms of tensionless oscillators. But even for any constant finite accelereation, we can see that action of the operator $(\L_0^{(\a)}-\bL_0^{(\a)})$ on a state remains invariant throughout this `flow' since $\Lambda_+^2-\Lambda_-^2=1$ by definition of Bogoliubov transformations, making sure this operator simply counts the difference in number of excitations in the two wedges.

\medskip

The above thus shows that the \textit{level-matched string states on the accelerated worldsheet remain level-matched throughout the flow in acceleration} as well, i.e all quantum closed string states clearly remains ``closed'' until actually hitting the infinite acceleration limit and collapsing on the horizon. This special symmetry ensures that despite being seemingly causally disconnected, excitations on either wedge knows about the other. 

\medskip

\paragraph{An aside on Hamiltonians:} It is intriguing to consider the Hamiltonian operator at finite acceleration. By the bogoliubov transformations mentioned earlier, this becomes
\be{hami2}
\L_0^{(a)}+\bL_0^{(a)}=\sum_m\left[ ( \Lambda_+^2+ \Lambda_-^2)( \b_{-m} \cdot \b_{m}+\tilde{\b}_{-m}\cdot \tilde{\b}_{m} )-4\Lambda_+ \Lambda_- \b_m\cdot\tilde{\b}_m    \right].
\ee
So the form of $\L_0^{(a)}+\bL_0^{(a)}$ combination does not remain invariant as we move to more and more accelerated strings, and the extra ``perturbation'' term generates a deformation, where the worldsheet physics differs from the inertial one.  One should remember, here we are just following the Minkowski space Hamiltonian as it changes with increasing acceleration, so evolution under this Hamiltonian is dissipative in nature due to the coupling term. Another novel thing to notice here is that, clearly oscillators $\b$ and $\tilde{\b}$ live in causally disconnected sectors, however they are here connected by the coupling term in the deformed Hamiltonian.  The curious implications of evolution under this hamiltonian would be discussed elsewhere. 

\medskip

\subsection{Gluing conditions, and coherent states}

It is clear from the above that even with the presence of certain special points, the embedding function on the accelerated ``closed'' worldsheet must be continuous everywhere, i.e. for a given string state $| B\rangle$, the Left wedge modes and the Right wedge modes (Left and Right parts of (\ref{rindlerws})) should smoothly go into one another
\be{bstate}
\left(X'^\mu_R (\eta,\xi)-X'^\mu_L (\tilde\eta,\xi)\right)_{\s=\pm\s_f}| B\rangle=0. 
\ee
Now as we described before, these two causally disconnected segments can be connected via an analytical continuation of the timelike coordinate on the two dimensional surface, which helps one to build global modes over the whole surface, i.e. one needs to consider $\tilde\eta=\eta+\frac{i\pi}{a}$ as we discussed in detail around (\ref{globmode}). This analytic continuation has to be taken care of when ``gluing'' the two halves of the string together. 

\medskip

Putting in the Left/Right wedge mode expansions from  (\ref{rindlerws}) into (\ref{bstate}) at $\eta=0$ we get the following conditions:
\bea{glue}
\left(\b_n^{\mu(R)}+ e^\frac{-\pi \lambda_n}{a} \tilde\b_{-n}^{\mu(L)} \right)| B\rangle&=&0,~~n\geq0\\
\left(q_0^{\mu(R)} -\tilde{q}_0^{\mu(L)}\right)| B\rangle&=&0
\eea
Here to arrive at the second equation, we have put separate centre of mass modes in the the two segments, for illustrative purposes. Now here we can take $| B\rangle$ to be some excited state defined in both wedges although they are causally disconnected.  The solution of the above conditions gives an expression  for such a state,      
\be{}
| B\rangle = \mathcal{N}\delta\left(q_0^{\mu(R)} -\tilde{q}_0^{\mu(L)}\right) \text{exp}\left[-\sum_m \frac{1}{m} e^\frac{-\pi \lambda_m}{a} \b_{-m}^{(R)}\cdot\tilde{\b}_{-m}^{(L)} \right]|0\rangle_{R}\otimes |0\rangle_{L},
\ee
which is a coherent state over the direct product vacuum that mixes degrees of freedom from both wedges. $\mathcal{N}$ is a normalisation factor, and the delta function specifies the position of the state. More specifically this is an entangled string state that couples the excitations in the two wedges.
\medskip

Now we can clearly compare the gluing condition (\ref{glue}) to the tensionless and tensile vacua evolution in (\ref{bcond}) by comparing the oscilators $(\b,\tilde{\b})$ and $(C,\C)$ as before and also identifying
\be{}
\tanh\theta =  \lim_{\e\to 0 }~\frac{\e-1}{\e+1} = - \lim_{a\to \infty }~e^\frac{-\pi \lambda_n}{a},
\ee
which has the right asymptotics near $a\to\infty$. At the point where the acceleration touches infinity, or equivalently at the Tensionless limit, one could see this coherent state evolves into 
\be{}
| B\rangle \sim\mathcal{\tilde{N}} \text{exp}\left[-\sum_m  \frac{1}{m} C_{-m}\cdot\tilde{C}_{-m} \right]|0\rangle_{R}\otimes |0\rangle_{L}.
\ee
This is a Neumann boundary condition in all spacetime directions{\footnote{Bear in mind, this is the limit where all points on the string collapses onto the $\s = \pm \t$ lines.}}. One could interpret this state as a $D_{p-1}$ brane i.e. a space filling brane. This is the same state we uncovered in (\ref{za2zc}) and signals appearance of open string degrees of freedom from closed strings as tension is dialed to zero. This takes us to interpret this state $| B\rangle $ as the inertial vacuum $|0\>_\a$ as seen by the observers associated to accelerated strings at the $a\to \infty$ limit. 

\medskip

The above ties in very well with our earlier observations about the what we called {\em{null string complimentarity}} in \cite{Bagchi:2020ats}. The idea there was that the two different sets of observers, ones on the inertial worldsheet observing the accelerated worldsheet and the ones on the accelerated worldsheet observing the inertial one, are related by the worldsheet Bogoliubov transformation. As the string turns null, the two different sets of observers have access to complementary pictures of open string physics emerging from closed strings. In our description of the gluing conditions in this section, we have been able to see that from the accelerated worldsheet the inertial closed string vacuum becomes a longer and longer string finally filling up the entire spacetime to form a spacefilling D-brane. We think that it is rather instructive that we can arrive at this picture from gluing conditions on the string worldsheet. 

\bigskip

\newpage

\section{Discussions and Conclusions}\label{sec8}

\subsection{Summary of results}
In this paper, we have laid the groundwork to addressing the important question of what happens when we consider strings near spacetime singularities. We have first shown that when strings probe black hole near horizons, this induces a tensionless limit on the string worldsheet. The near horizon limit is a Carrollian limit in spacetime and hence this Carroll limit induces a Carroll limit on the worldsheet as well. 

\medskip

We then turned our attention to strings propagating in $d$ dimensional Rindler spacetimes. This is a precursor to the more involved question of strings propagating in the full near horizon geometry which consists of a $2d$ Rindler and additional transverse dimensions. As we emphasised earlier, the main results of our analysis in the rest of the paper is expected to hold when considering this more general question. 

\medskip

We argued that the background Rindler spacetime induced a Rindler structure on the string worldsheet and this led to novelties appearing on the worldsheet, in terms of foldings and changing periodicities. The foldings are brought about here by the acceleration of the worldsheet and are reminiscent of folded spinning strings in flat or $AdS$ spacetimes. When studying periodicity, we found that this rapidly moves away from $2\pi$ as one approaches the worldsheet horizon, indicating the presence of special points on the worldsheet which divide it into causally disconnected segments. These novelties depend on the appearance of a horizon on the worldsheet induced from the background horizon, a feature that would continue to persist when we consider strings moving in the full near horizon geometry.   

\medskip

As the horizon on the Rindler worldsheet was approached and finally hit, we saw the emergence of BMS symmetries from the usual Virasoro symmetries of the closed worldsheet.  Here we made the interesting observation of the formation of Rindler closed strings in both wedges from different segments and its link with two different ways of contracting the Virasoro algebra to BMS. 

\medskip

Finally, in Sec \ref{sec7}, we had a detailed discussion of the quantum states of the accelerated string. We performed mode expansions, linked this to our picture of string segments and understood change of string vacua via Bogoliubov transformations. We addressed the question of the evolution of the level matching conditions under acceleration and showed that the closed string remains closed until the very end of the flow when the horizon is hit. (Even at the end, it retains some memory of this closure when it turns into an open string.) We used this property to define gluing conditions at the folding points of the string and showed the emergence of the boundary state and its link to null string complementarity. 

\medskip

In a nutshell, one of the main things we have shown in this paper is that the worldsheet methods advocated in \cite{Bagchi:2020ats} can be given a firm footing in terms of strings propagating in background Rindler spacetimes. Our analysis is expected to play a pivotal role in understanding strings near generic spacetime singularities.  We should emphasize however, that we are attempting to understand physics near and on the horizon but have nothing yet to say for going beyond the horizon. This is obviously an important question for future studies. 

\subsection{Future directions}

There are some immediate directions of research that follow from our considerations in the current paper, and there are some other tantalising hints which merit some discussion. 

\medskip

As we have repeatedly stressed throughout the paper, we consider our work in this paper an important step in the direction of a more involved problem, viz. understanding strings propagating near generic black hole horizons. As explained in the introduction, quantizations of strings in the near horizon region is likely to involve a different mathematical machinery in the form of Newton-Cartan structures. Issues with the transverse spherical directions, which we have carefully avoided in this paper, would play a crucial role there. This is work in progress and we hope to report about this sometime in the near future. 

\medskip

In Sec.~\ref{7.2}, we have briefly mentioned the evolution of the Hamiltonian of the string along the flow of acceleration (\ref{hami2}). This can be connected concretely to the generation of entanglement between two `halves' of the string as it accelerates.  We will report on the evolution of this Hamiltonian and its links with entanglement on the worldsheet in upcoming work. 

\medskip

We have explained at the beginning of this paper that we need to distinguish between the evolution of closed string theory in boosts and its evolution in acceleration. The evolution in boosts does not change the physics until the boost becomes infinite and thereby making the string null. The evolution in acceleration, however, changes the vacuum all along the flow and this is given by the Bogoliubov transformations on the worldsheet and is the unique new physics we have tried to explain in this work. The limit of infinite acceleration also lands us on the null string and hence physics described in this limit is impervious to whether we look at infinite boosts or infinite accelerations. It would be instructive to understand these evolutions directly in terms of boosted and accelerated strings and figure out the emergence of BMS from the Virasoro algebra in this way. This again is work in progress. 

\medskip

Last but not the least, the supersymmetric versions of our tensionless strings have been constructed in \cite{Bagchi:2016yyf, Bagchi:2017cte}. For these, the worldsheet symmetry algebras are different avatars of Super BMS algebras and they have much richer structures than purely bosonic ones. It would be very interesting to understand the Rindler picture advocated in this work in the supersymmetric context. We hope to return to this problem in the near future.

\subsection{Discussions} 

We finally end with some comments about existing literature and possible connections to our work. 

\medskip

A nice spacetime picture of free strings falling onto a black hole horizon, where the near horizon is approximated by a Rindler space, was advocated by Susskind and collaborators \cite{Susskind:1993if, Susskind:1994sm, Susskind:1994uu}. In this picture a stationary bulk Rindler observer looking at a free falling string in large fall times, or under large Lorentz boosts, will see the transverse size of the string grow very fast as it hits the horizon. The authors conclude that infalling strings near the horizon pass onto the long string phase and diffuse over the horizon at a distance of order $\sqrt{\a'}\sim l_s$, creating what is known as a `stretched horizon'. The stretched horizon is an effective membrane that has all the qualitative features of the horizon it encapsulates. This `soup of strings' are said to be in thermal equilibrium at the temperature $\sim \frac{1}{\sqrt{\a'}}$, also known as the Hagedorn temperature, which incidentally is where tensionless strings are also supposed to emerge according to the lore in literature  \cite{Pisarski:1982cn}.  

\medskip

Although this picture seemingly resonates with our current discussion, there are important differences which we cannot immediately resolve. In our analysis the $\a'$ cannot be kept fixed, as is the essence of the tensionless limit. Hence we don't have a notion for the minimal length scale corresponding to  $\sqrt{\a'}\sim l_s$, which for us would correspond to a limiting or maximal acceleration. It is clearly seen in our computations that BMS$_3$ as a worldsheet algebra is generated only when the string completely hits the horizon, or at the strict $a\to \infty$ limit.
It would be nice to clarify this point of departure between the approaches.

\medskip

We have spoken in details about the natural appearance of folds on the worldsheet due to the induced worldsheet horizon in our setup. In recent works e.g. \cite{Giveon:2019twx}, folded strings appear in a completely different setting, that of a realisation of the ER=EPR in the string context. The authors arguments rely on entangled folded strings on eternal two-sided black hole geometry and corresponding target space interpretation of CFT fusion. They suggest that in the Lorentzian picture, the analytically continued folded string lives simultaneously in two disconnected flat spaces, with one copy each in two, entangling the space-times. More recently, in \cite{Jafferis:2021ywg} this idea has been taken further and Lorentzian continuation of winding strings have been identified as pairs of entangled folded strings emanating from the horizon. These strings ending on the horizon have been conjectured to be the hot open string soup near horizons which make up for the BH entropy in string theory \cite{Susskind:1994sm}. A lot of the discussion in these works at first sight seem to be related to the picture we have tried to build in this paper. A concrete connection is of course lacking at present, but we hope in future work, we would be able to provide more evidence and a bridge between these very interesting discussions and our tensionless strings point of view.


\medskip


\medskip


\bigskip 

\bigskip

\newpage

\section*{Acknowledgements}
We would like to thank Joan Simon for collaboration on the initial parts of the paper and also for many detailed discussions and probing questions which we believe has helped increase our understanding of the problem at hand.  

\medskip

AB is partially supported by a Swarnajayanti fellowship of the Department of Science and Technology, India and by the following grants from the Science and Engineering Research Board: SB/SJF/2019-20/08, MTR/2017/000740, CGR/2020/002035. 
The work of ArB is supported by the Quantum Gravity Unit of the Okinawa Institute of Science and Technology Graduate University (OIST). ArB would like to thank APCTP Pohang and Kyoto University for hospitality at various stages of this work being done. 
SC is partially supported by the ISIRD grant 9-252/2016/IITRPR/708. RC is supported by the CSIR grant File No: 09/092(0991)/2018-EMR-I.

\bigskip
\bigskip
\bigskip

\section*{APPENDIX}

\appendix

\section{Orthonormality of Rindler modes} \label{ApA}

The mode expansion of the closed bosonic string in an inertial frame is 

\begin{equation}
    X^{\mu}(\sigma,\tau)=x^{\mu}+ \alpha' p^{\mu}\tau+\sqrt{2\pi\alpha'}\sum_{n> 0}[\alpha^{\mu}_{n}u_{n}+\alpha^{\mu}_{-n}u^{*}_{n}+\Tilde{\alpha}^{\mu}_{n}\Tilde{u}_{n}
    +\Tilde{\alpha}^{\mu}_{-n}\Tilde{u}^{*}_{n}], 
    \label{minkws}
\end{equation}
where the Right and Left moving modes are given by 
\be{}
u_{n}=\frac{ie^{-in(\tau-\sigma)}}{\sqrt{4\pi}n},~~ \Tilde{u}_{n}=\frac{ie^{-in(\tau+\sigma)}}{\sqrt{4\pi}n}.
\ee 
From the reality of mode expansion we get  $\a^\dagger_{n} = \a_{-n}$. The worldsheet vacuum is defined as, 
\be{alphaa}
\a_n|0\rangle_{\a}= \tilde\a_n|0\rangle_{\a}=0.
\ee
Mode expansion of the closed string on entire Rindler  worldsheet comprising left and right wedges is, 
\begin{equation} 
    X'^{\mu}(\xi,\eta)=q^{\mu}+\alpha'p^{\mu}\eta+\sqrt{2\pi\alpha'} \sum_{n> 0}[\beta_{n}U_{n}{(R)}+\tilde\beta_{n} \tilde{{U}}_{n}(L) +h.c.].
\end{equation}


By using the relation between inertial worldsheet coordinate and Rindler worldsheet coordinates,
\begin{align}
 &&    \sigma+\delta = e^{a(\xi-\chi)}\cosh{a\eta},~ \tau = e^{a(\xi-\chi)}\sinh{a\eta} ~~~ \text{(\textit{Right} wedge)}\nonumber \\
 &&       \sigma+\delta = -e^{a(\xi-\chi)}\cosh{a\eta}, ~\tau = -e^{a(\xi-\chi)}\sinh{a\eta} ~~~ \text{(\textit{Left} wedge)} 
\end{align}
we present various positive frequency modes on Rindler worldsheet, 
\begin{eqnarray}
U_n(R)=\left\{
\begin{array}{l}
\, U_n^{(R)} =  \frac{ie^{-i\lambda_{n}(\xi-\eta)}}{\sqrt{4\pi}n} =   \frac{i}{\sqrt{4\pi}n} e^{-i \lambda_n \chi} {(\sigma -\tau +\delta)}^{\frac{- i \lambda_n}{a}}
,\\\\ \nonumber\\
U_{-n}^{(R)} =  \frac{-ie^{i\lambda_{n}(\xi+\eta)}}{\sqrt{4\pi}n} =   \frac{-i}{\sqrt{4\pi}n} e^{i\lambda_n \chi} {(\sigma +\tau +\delta)}^{\frac{ i \lambda_n}{a}}  \,,\\
\end{array}\right.\\\\\nonumber\\
\tilde{U}_n(L)=\left\{
\begin{array}{l}
\, {{U}}_n^{(L)} =  \frac{ie^{-i\lambda_{n}(\xi+\eta)}}{\sqrt{4\pi}n} =   \frac{i}{\sqrt{4\pi}n} {(-1)}^{\frac{-i \lambda_n}{a}}e^{-i \lambda_n \chi} {(\sigma + \tau +\delta)}^{\frac{- i \lambda_n}{a}}
,\\\\
{{U}}_{-n}^{(L)} =  \frac{-ie^{i\lambda_{n}(\xi-\eta)}}{\sqrt{4\pi}n} =   \frac{-i}{\sqrt{4\pi}n} {(-1)}^{\frac{i \lambda_n}{a}}e^{i \lambda_n \chi} {(\sigma - \tau +\delta)}^{\frac{i \lambda_n}{a}}
\\
\end{array}
\right.
\end{eqnarray}

Likewise the negative frequency modes are,
\begin{eqnarray}
U_n(R)^*=\left\{
\begin{array}{l}
\,  U_n^{(R)*} =  \frac{-ie^{i\lambda_{n}(\xi-\eta)}}{\sqrt{4\pi}n} =   \frac{-i}{\sqrt{4\pi}n} e^{i \lambda_n \chi} {(\sigma -\tau +\delta)}^{\frac{ i \lambda_n}{a}}
,\\\\ \nonumber\\
  U_{-n}^{(R)*} =  \frac{ie^{-i\lambda_{n}(\xi+\eta)}}{\sqrt{4\pi}n} =   \frac{i}{\sqrt{4\pi}n} e^{-i\lambda_n \chi} {(\sigma +\tau +\delta)}^{\frac{- i \lambda_n}{a}} \,,\\
\end{array}\right.\\\\\nonumber\\
\tilde{U}_n(L)^*=\left\{
\begin{array}{l}
\,   
  {{U}}_n^{(L)*} =  \frac{-ie^{i\lambda_{n}(\xi+\eta)}}{\sqrt{4\pi}n} =   \frac{-i}{\sqrt{4\pi}n} {(-1)}^{\frac{i \lambda_n}{a}}e^{i \lambda_n \chi} {(\sigma + \tau +\delta)}^{\frac{ i \lambda_n}{a}} ,\\\\
 {{U}}_{-n}^{(L)*} =  \frac{ie^{-i\lambda_{n}(\xi-\eta)}}{\sqrt{4\pi}n} =   \frac{i}{\sqrt{4\pi}n} {(-1)}^{\frac{-i \lambda_n}{a}}e^{-i \lambda_n \chi} {(\sigma - \tau +\delta)}^{\frac{-i \lambda_n}{a}}\\
\end{array}
\right.
\end{eqnarray}

We define the inner product in the right wedge of the Rindler worldsheet. 
\begin{align}
(\phi_1(R),\phi_2(R)) =  i \int d\xi \Big[ \phi_1(R)\partial_{\eta}\phi_2(R)^*  - \phi_2(R)^* 
\partial_{\eta} \phi_1(R) \Big]
\end{align}

Similarly, in the left wedge of Rindler worldsheet the definition of inner product goes as, 
\begin{align}
(\phi_1(L),\phi_2(L)) =  - i \int d\xi \Big[ \phi_1(L)\partial_{\eta}\phi_2(L)^*  - \phi_2(L)^* \partial_{\eta} \phi_1(L) \Big]
\end{align}
With these definitions of inner product we check the orthonormality of local modes defined in the individual wedges of Rindler worldsheet. In order to check that, we redefine respective local modes $V_n^{(R)} = \sqrt{n} ~ U_n^{(R)}$ and $V_n^{(L)} = \sqrt{n} ~ U_n^{(L)}$ for our computational convenience. In the right Rindler wedge,
\begin{align}
(V_m^{(R)}, V_n^{(R)}) &= i \int_0^{L_\delta} d\xi \Big[ V_m^{(R)} \partial_{\eta} V_n^{(R)*} -  V_n^{(R)*} \partial_{\eta} V_m^{(R)}\Big] \nonumber \\
&= i \int_0^{L_\delta} d\xi \Big[ \Big( \frac{i e^{-i\lambda_m(\xi-\eta)}}{\sqrt{4 \pi m}}  \Big) \partial_{\eta} \Big(\frac{-i e^{i\lambda_n(\xi-\eta)}}{\sqrt{4 \pi n}}  \Big) -  \Big(\frac{-i e^{i\lambda_n(\xi-\eta)}}{\sqrt{4 \pi n}}  \Big) \partial_\eta \Big( \frac{i e^{-i\lambda_m(\xi-\eta)}}{\sqrt{4 \pi m}}  \Big) \Big]\nonumber \\
&= \frac{1}{4 \pi \sqrt{mn}}   \int_0^{L_\delta} d\xi \Big[ \Big( \frac{2 \pi n}{L_\delta}\Big)+\Big(\frac{2 \pi m}{L_\delta}\Big) \Big]e^{\frac{- 2\pi }{L_\delta} (m-n)(\xi-\eta)}\nonumber \\
&= \delta_{m,n}
\end{align}

Further we can show, 
\begin{eqnarray}
&&( {V}_{-m}^{(R)}, {V}_{-n}^{(R)})=\delta_{m,n}  \nonumber \\
&&(V_m^{(R)*}, V_n^{(R)*}) = ( {V}_{-m}^{(R)*}, {V}_{-n}^{(R)*}) = -\delta_{m,n} \nonumber \\
&&(V_m^{(R)}, V_n^{(R)*}) = (V_m^{(R)},V_{-n}^{(R)*} )= 0
\end{eqnarray}
For left Rindler wedge, using the appropriate definition of inner product one can check the orthonormality of local modes in a straightforward way. We elaborate upon one example and mention the final results for others. 
\begin{align}
(V_m^{(L)}, V_n^{(L)}) = &= -i \int_0^{L_\delta} d\xi \Big[ V_m^{(L)} \partial_{\eta} V_n^{(L)*} -  V_n^{(L)*} \partial_{\eta} V_m^{(L)}\Big] \nonumber \\
&= -i \int_0^{L_\delta} d\xi \Big[ \Big( \frac{i e^{-i\lambda_m(\xi+\eta)}}{\sqrt{4 \pi m}}  \Big) \partial_{\eta} \Big(\frac{-i e^{i\lambda_n(\xi+\eta)}}{\sqrt{4 \pi n}}  \Big) -  \Big(\frac{-i e^{i\lambda_n(\xi+\eta)}}{\sqrt{4 \pi n}}  \Big) \partial_\eta \Big( \frac{i e^{-i\lambda_m(\xi+\eta)}}{\sqrt{4 \pi m}}  \Big) \Big]\nonumber \\
&= \frac{1}{4 \pi \sqrt{mn}}   \int_0^{L_\delta} d\xi \Big[ \Big( \frac{2 \pi n}{L_\delta}\Big)+\Big(\frac{2 \pi m}{L_\delta}\Big) \Big]e^{\frac{- 2\pi }{L_\delta} (m-n)(\xi-\eta)}\nonumber \\
&= \delta_{m,n}
\end{align}
\begin{eqnarray}
&&  ( {V}_{-m}^{(L)}, {V}_{-n}^{(L)}) = \delta_{m,n} \nonumber \\
&& (V_m^{(L)*}, V_n^{(L)*}) = ( {V}_{-m}^{(L)*}, {V}_{-n}^{(L)*}) = -\delta_{m,n} \nonumber \\
&& (V_m^{(L)}, V_n^{(L)*}) = (V_m^{(L)},{V}_{-n}^{(L)*}) = 0
\end{eqnarray}
The mode expansion of closed string in terms of Rindler global mode is as  follows, 
 \begin{equation}
    X^{\mu}(\xi,\eta)=q_{RG}^{\mu}+\alpha'p_{RG}^{\mu}\eta+\sqrt{2\pi\alpha'} \sum_{n> 0}[\gamma^{\mu}_{n}h_{n}+\gamma^{\mu}_{-n}h^{*}_{n}+\Tilde{\gamma}^{\mu}_{n}\Tilde{h}_{n}+\Tilde{\gamma}^{\mu}_{-n}\Tilde{h}^{*}_{n}],
 \end{equation}
 where the global modes $h_n = \frac{H_n}{\sqrt{n}}$ and $\tilde{h}_n = \frac{\tilde{H}_n}{\sqrt{n}} $ are defined as, 
\begin{align}
H_{n}&=\mathcal{N}_1\Big( V_{n}^{(R)}-{(-1)}^{i \frac{\lambda_{n}}{a}}V^{(L)*}_{-n}\Big)\nonumber \\
\tilde{H}_{n}&=\mathcal{N}_2 \Big(V_{n}^{(L)}-{(-1)}^{- i \frac{\lambda_{n}}{a}}V^{(R)*}_{-n}\Big).
\end{align}
$\mathcal{N}_1$ and $\mathcal{N}_2$ are introduced as the normalization constants. 
We demand the orthonormality of the Rindler global modes, 
\begin{eqnarray}
&(H_{n}, H_{n}) =1 \\
\label{glob1}
& (\tilde{H}_{n}, \tilde{H}_{n}) = 1
 \label{glob2}
\end{eqnarray}
As a consequence of demanding Rindler global mode to be orthonormal leads to the evaluation of $\mathcal{N}_1$.
 \begin{eqnarray}
&&|\mathcal{N}_1|^2 \Big[ (V_{-n}^R, V_{-n}^R) + {(-1)}^{\frac{2 i \lambda_n}{a}} (V_n^{L *},V_n^{L *}) \Big] = |\mathcal{N}_1|^2  \Big[ 1 - e^{\frac{- 2 \pi \lambda_n}{a}}  \Big] =1 \nonumber \\
&& |\mathcal{N}_1|^2 e^{\frac{-  \pi \lambda_n}{a}} \Big[ e^{\frac{ \pi \lambda_n}{a}} - e^{\frac{- \pi  \lambda_n}{a}}  \Big] = |\mathcal{N}_1|^2 e^{\frac{- \pi  \lambda_n}{a}} 2 \sinh\Big(\frac{\pi \lambda_n}{a}\Big) = 1\nonumber \\
&& |\mathcal{N}_1| = \frac{e^{\frac{ \pi \lambda_n}{2a}}}{\sqrt{2 \sinh\Big(\frac{\pi \lambda_n}{a}\Big) }}
\end{eqnarray}
Note that we use the relation $(-1) = e^{i \pi}$ in the above derivation of $\mathcal{N}_1$.

Following a similar analysis, we can also fix $\mathcal{N}_2$. However, in order to do that, we use the relation $(-1) = e^{-i \pi}$.
\begin{eqnarray}
 |\mathcal{N}_2| = \frac{e^{\frac{ \pi \lambda_n}{2a}}}{\sqrt{2 \sinh\Big(\frac{\pi \lambda_n}{a}\Big) }}
\end{eqnarray}

By taking account of the two normalization constants it is immediate to figure out the desired orthonormal Rindler global modes, 
\begin{align}
h_{n}&=\frac{1}{\sqrt{2\sinh{\Big(\frac{\pi\lambda_{n}}{2a}\Big)}}}\Big(e^{\frac{\pi\lambda_{n}}{2a}}U_{n}^{(R)}-e^{-\frac{\pi\lambda_{n}}{2a}}{U}^{(L)*}_{-n}\Big)\nonumber \\
\Tilde{h}_{n}&=\frac{1}{\sqrt{2\sinh{\Big(\frac{\pi\lambda_{n}}{2a}\Big)}}}\Big(e^{\frac{\pi\lambda_{n}}{2a}}{U}^{(L)}_{n}-e^{-\frac{\pi\lambda_{n}}{2a}}U^{(R)*}_{-n}\Big).
\end{align}

\newpage

\newpage

\bibliographystyle{JHEP}
\bibliography{ref}

\end{document}